\documentclass[pra, twocolumn, superscriptaddress, nofootinbib]{revtex4-2}

\usepackage{graphicx}
\usepackage{amsmath}
\usepackage{amssymb}
\usepackage{braket}
\usepackage{hyperref}
\usepackage{cleveref}
\usepackage{amsthm}

\usepackage{mathtools}
\usepackage{physics}
\usepackage{algorithm}
\usepackage{algpseudocode}
\usepackage{placeins}
\begin{document}
	
	\title{Quantum State Engineering Under Multiple Expectation-Value Constraints}
	
	\author{Anjali Mahapatra}
	\author{Gururaj Kadiri}
	\email{gururaj@igcar.gov.in}
	
	\affiliation{Materials Science Group, Indira Gandhi Centre for Atomic Research, Kalpakkam 603102, Tamil Nadu, India}
	\affiliation{Homi Bhabha National Institute, Training School Complex, Anushaktinagar, Mumbai 400094, India}
	
	\date{\today}
	
	
	\begin{abstract}
		
		This work introduces a formulation of quantum state engineering termed  expectation-value targeting: the task of preparing a pure state whose expectation values with respect to a prescribed set of observables attain specified targets. This formulation subsumes standard ground-state preparation problems in quantum chemistry and many-body physics, while extending beyond variational energy minimization to multi-constraint state synthesis. The problem amounts to solving a system of nonlinear constraints on an exponentially large state space, for which no general efficient classical approaches are known. Variational quantum algorithms tackle this problem by restricting the search to a low-dimensional parameter space, and relying on classical optimization techniques for solutions. However, these approaches can become extremely ineffective for the present problem, where competing constraints can induce rugged landscapes and vanishing gradients (barren plateaus).
		Adaptive variational methods, in which the ansatz is constructed iteratively from a pool of candidate operators rather than fixed in advance, have been developed primarily for ground-state preparation. However, we show that the present problem also admits a similar construction. We introduce QUEST (Quantum Unitary Engineering of States to Target), a framework purpose-built for expectation-value targeting, in which the engineered state is constructed as a depth-adaptive sequence of Pauli rotations, with each rotation chosen to descend a sum-of-squared-residuals cost. QUEST provides a constructive route to expectation-value targeting, building the engineered state one Pauli rotation at a time, and establishes adaptive synthesis as a primitive for state preparation under multiple, potentially inconsistent target constraints.
	\end{abstract}
	\maketitle
	
	\section{Introduction}
	Preparing quantum states with prescribed physical properties is a central task in quantum simulation\cite{zhang2022quantum,rosenkranz2025quantum,cao2019quantum}, control, and metrology \cite{yang2024quantum,zuniga2024variational,giovannetti2004quantum}. In many settings, these properties are specified not by a wavefunction or circuit description, but by the expectation values of a set of observables\cite{brif2010control,mcclean2016theory,khaneja2005optimal,giovannetti2004quantum,paris2009quantum,bravyi2002fermionic,cramer2010efficient,bravyi2017tapering,calabrese2006time,wiseman1993quantum}. This leads to a natural formulation of quantum state engineering under expectation-value constraints: given operators $\{O_i\}$ and target values $\{\tau_i\}$, construct a pure state $|\psi\rangle$ such that $\langle \psi | O_i | \psi \rangle = \tau_i$ for all $i$, under the promise that such a state exists.  A closely related feasibility question, whether a given set of reduced density matrices is consistent with a global quantum state, has been studied extensively as the quantum marginal problem \cite{coleman1963structure,klyachko2004quantum,schilling2014quantum}, and as N-representability in quantum chemistry \cite{smith1965n,klyachko2006quantum}. Expectation-value targeting generalizes this setting by allowing arbitrary (not necessarily local, not necessarily complete) operator sets, and focuses on the constructive preparation of a pure state rather than on characterizing the feasible region. This formulation encompasses standard ground-state preparation as a special case, with a single constraint $\langle H \rangle = E_0$, but extends more broadly to multi-constraint settings where physically relevant states are defined by simultaneous conditions on energy, correlations, symmetries, or witness operators. Unlike variational formulations \cite{cerezo2021variational} that reduce the task to minimizing a single objective, expectation-value constraints define a feasibility problem on the quantum state space, where multiple conditions must be satisfied simultaneously and may compete with one another. Making this structure explicit exposes a class of state-engineering problems that is implicit in many applications but has not been treated as a primary object of study.
	
	Despite its generality, expectation-value-constrained state engineering is challenging. A direct classical approach requires representing the quantum state explicitly, leading to a description that grows exponentially with system size and reduces the task to solving a system of nonlinear constraints on a high-dimensional manifold. Variational quantum algorithms \cite{mcclean2016theory} address this by parameterizing the state with a quantum circuit and optimizing a cost function that penalizes deviations from the target values. While this reduces the dimensionality of the search, it introduces a classical optimization loop over a non-convex landscape. In the multi-constraint setting, competing objectives can lead to flat directions and weak optimization signals, making convergence particularly difficult. These limitations motivate the development of approaches that avoid high-dimensional classical optimization while directly exploiting the structure of expectation-value constraints.
	We present QUEST (Quantum Unitary Engineering of States to Target), a constructive, optimizer-free quantum algorithm that solves expectation-value targeting by adaptively building a Pauli path (also called an adaptive ansatz in the ADAPT-VQE literature) as an ordered product of Pauli rotations with all parameters determined analytically at each step. The analytical tractability rests on a single structural fact: the multi-constraint cost is an exact five-parameter trigonometric polynomial in any single rotation angle, fully characterized by five oracle queries and minimized in closed form, eliminating the need for a classical optimization loop entirely. \par
	This work serves two purposes. First, we formalize expectation-value targeting as a distinct class of quantum state engineering problems. Second, we introduce QUEST as an algorithmic framework for this class. 
	The rest of the paper is arranged as follows: In Section \ref{sec:background} we develop the mathematical framework of the problem of expectation value targetting and explain how the conventional ADAPT-VQE can be modified to solve this problem, and then discuss the shortcomings of this approach. In Section \ref{sec:Framework} we introduce the QUEST algorithm for solving the expectation-value targetting problem, and discuss its features. 
	Section \ref{sec:Results} is devoted to examining the performance of QUEST in comparision with the ADAPT-VQE in solving the expectation-value targetting problems posed in three different quantum domains. Section \ref{sec:Conclusion} concludes the manuscript by discussing the scope and future prospects of this work. 
	\section{Background}
	\label{sec:background}
	Let $\mathcal{O}$ be a quantum oracle that returns the expectation value $\bra{\psi}\hat{O}\ket{\psi}$ for any prepared state $\ket{\psi}$, with $\hat{O}$ belonging to a set of Hermitian operators $\{\hat{O}_i, \quad i=1,\dots,N\}$. 
	The problem of multi-constraint state engineering is to prepare an n-qubit quantum state $\ket{\phi}$ such that 
	\begin{equation}
		\bra{\phi}\hat{O}_i\ket{\phi}=  \tau_i, \quad i=1,\dots,N,
		\label{eq:target_equation}
	\end{equation}
	under the promise that at least one pure state satisfying all $N$ constraints of Eq. (\ref{eq:target_equation}) exists.
	A direct classical approach requires an explicit representation of the state,
	$|\psi\rangle = \sum_{x=0}^{2^n-1} c_x |x\rangle$,
	involving $O(2^n)$ real parameters (up to normalization and global phase). The problem then reduces to minimizing $C(\psi) = \sum_i \big( f_i(\psi) - \tau_i \big)^2$ over this exponentially large parameter space using only oracle evaluations. This constitutes a system of nonlinear constraints on $\mathbb{CP}^{2^n-1}$ with no accessible structure beyond function queries, and no efficient classical method is known in general. 
	In the variational hybrid methods \cite{peruzzo2014variational,tilly2022variational}, starting from an initial state $\ket{\chi}$, the target state is expressed through a parameterized circuit $\hat{U}(\vec{\theta})$ as:
	\begin{equation}
		\ket{\psi(\vec{\theta})}=\hat{U}(\vec{\theta})\ket{\chi}
	\end{equation}
	where $\vec{\theta}$ is a collection of real parameters, representing the parameters of an arbitrary unitary transformation, which, for instance,  could be a hardware-efficient ansatz \cite{kandala2017hardware,sim2019expressibility} composed of native gate sets (CNOT, $R_z$, $R_x$), or a chemically motivated UCCSD \cite{bartlett2007coupled,yung2014transistor,taube2006new,harsha2018difference,kutzelnigg1991error} ansatz. Although typically employed for minimizing the Hamiltonian, VQE can also be employed for our problem, by defining a multi-constraint cost-function which has to be minimized:
	\begin{equation}
		C(\vec{\theta}) = \sum_{i=1}^N \left( \langle \psi(\vec{\theta}) | O_i | \psi(\vec{\theta}) \rangle - \tau_i \right)^2.
		\label{eq:vqe_cost_target}
	\end{equation}
	Although the cost function in Eq. (\ref{eq:vqe_cost_target}) involves a sum of multiple terms, the present problem is not a multi-objective optimization \cite{ekstrom2025variational} which seek Pareto-optimal trade-offs among competing objectives. Multi-objective optimization is typically formulated as an optimization problem over classical parameters, even when the objective functions are evaluated using quantum resources. In contrast, the present formulation is intrinsically quantum, as it concerns the existence and construction of quantum states satisfying a set of expectation-value constraints.
	
	In contrast to the standard VQE, where the Hamiltonian $H$ is known explicitly, here we only have access to an Oracle $\mathcal{O}$. The expectation values appearing on the RHS of Eq. (\ref{eq:vqe_cost_target}) are estimated via repeated quantum measurements using this Oracle. The parameters $\vec{\theta}$ are optimized by classical optimizers like gradient descent, L-BFGS \cite{liu1989limited}, or SPSA \cite{spall1998overview,spall1997accelerated,kandala2017hardware}, which iteratively update $\vec{\theta}$ to minimize the cost function $C(\vec{\theta})$. Gradients with respect to $\vec{\theta}$ are computed either analytically via the parameter-shift rule \cite{wierichs2022general} or by finite differences, with each gradient evaluation requiring a number of circuit executions proportional to the number of parameters. The central shortcoming of the VQE family lies in this classical optimization loop, which displays a well-known pathological feature called the barren plateaus \cite{larocca2025barren,wang2021noise,mcclean2018barren,cunningham2025investigating}. These are regions on the energy landscape of exponentially  vanishing gradients, rendering the angle updates  ineffective. Even away from barren plateaus, the non-convex landscape harbors local minima and saddle points, and convergence to the global minimum is not guaranteed. 
	Unlike the Hamiltonian expectation value, the cost $C(\vec{\theta})$ is a sum of multiple constraints. A parameter update that improves one expectation value may worsen another, leading to competing gradients and possible cancellation effects. Consequently, the gradient of $C(\vec{\theta})$ may become small even when the state $\ket{\psi(\vec{\theta})}$ is far from satisfying all the $N$ constraints. \par
	Furthermore, in conventional VQE applications, the structure of the Hamiltonian can be exploited to design tailored ans\"atze \cite{anselmetti2021local,burton2024accurate,gocho2023excited} or optimization strategies. The present problem, being cast in the Oracle model, prohibits this. The operators $\{O_i\}$ are not directly accessible, preventing any structure-aware optimizations, rendering the cost function effectively unstructured from the algorithm's perspective. 
	ADAPT-VQE[\cite{grimsley2019adaptive,tang2021qubit,feniou2023overlap,anastasiou2024tetris,yordanov2021qubit,ramoa2025reducing}] addresses the ansatz rigidity of conventional VQE by constructing the circuit adaptively. Starting from an initial state $|\chi\rangle$, it iteratively appends Pauli rotations  $e^{i\theta P}$ at the end of the circuit, with the Pauli operator $P$ selected from a predefined operator pool, chosen as the operator whose gradient $\partial\langle\hat{H}\rangle/\partial\theta|_{\theta=0}$ is largest in magnitude. The appended angle $\theta$ is then determined by re-optimizing the full parameter vector of the new circuit via a classical optimizer, as with VQE. \par
	While, in general, ADAPT-VQE produces more compact circuits than fixed-ansatz VQE and avoids committing to a circuit structure a priori, it also provides no guarantee of 
	convergence to the global minimum, leaving the algorithm susceptible to local minima and stagnation when the gradient signal is weak. These shortcomings are 
	further sharpened in the multi-constraint setting considered here.  We demonstrate two dramatic illustrations of this failure mode in Sec.~\ref{sec:Results},  where the gradient vanishes identically for every pool element from the outset, with no scope for ADAPT-VQE to progress even a single step. 
	\section{QUEST Framework}
	\label{sec:Framework}
	\subsection{Theoretical Background}
	We refer to the unitary transformation of the form $e^{i\theta P}$ as a ``Pauli rotation'', and read it as ``Pauli P oriented at the angle $\theta$''. The central object of interest in QUEST is what we call ``Pauli path", which is the ordered product of Pauli rotations:
	\begin{equation}
		\hat{U}(\vec{\theta},\vec{P}) = e^{i\theta_k P_k}\cdots e^{i\theta_1 P_1},
		\label{eq:pauli_path}
	\end{equation}
	where $\vec{P} = (P_1,\ldots,P_k)$ is a sequence of Pauli strings and $\vec{\theta} = (\theta_1,\ldots,\theta_k)$ the corresponding angles.
	We shall refer to the state obtained by acting $\hat{U}(\vec{\theta},\vec{P})$ on a state $|\chi\rangle$ as $\ket{\chi(\vec{\theta},\vec{P})}$: 
	\begin{equation}
		\ket{\chi(\vec{\theta},\vec{P})} = \hat{U}(\vec{\theta},\vec{P})|\chi\rangle,
		\label{eq:psi_theta_P}
	\end{equation}
	and denote  the expectation value of a Hermitian operator $\hat{O}$ in the terminal state $\ket{\chi(\vec{\theta},\vec{P})}$ by $O(\vec{\theta},\vec{P})$:
	\begin{equation}
		O(\vec{\theta},\vec{P}) = \bra{\chi(\vec{\theta},\vec{P})}\hat{O}\ket{\chi(\vec{\theta},\vec{P})}	\label{eq:operator_expectation_value}
	\end{equation}
	For a given initial state $\ket{\chi}$, we define the cost function, analogous to Eq. (\ref{eq:vqe_cost_target}) as 
	\begin{equation}
		C(\vec{\theta},\vec{P}) = \sum_{i=1}^N w_i\left(O_i(\vec{\theta},\vec{P}) - \tau_i\right)^2
		\label{eq:cost_function}
	\end{equation}
	\par
	Here $\textbf{w}=(w_1,w_2,...w_N)$ is the weight vector, where $w_i \ge 0$ are the weights associated with each constraint.
	
	The aim of QUEST to identify a Pauli path for which this cost function is $0$. \par
	Given a Pauli path $\hat{U}(\vec{\theta},\vec{P})$, QUEST can alter it by inserting a new Pauli rotation between the $l^{th}$ and $(l+1)^{th}$ Pauli rotations of $\hat{U}(\vec{\theta},\vec{P})$.
	We use the notation $\hat{U}(\vec{\theta},\vec{P};\,l;\,\theta,\,P)$ to denote the new Pauli-path,  with $l$ indicating the position of insertion, $\theta$ indicating the angle, and $P$ indicating the Pauli string. Let the corresponding terminal states be $\ket{\chi(\vec{\theta},\vec{P};\,l;\,\theta,\,P)}$. 
	The expectation values of an operator $\hat{O}$ in the corresponding terminal states are:
	\begin{equation}
		O(\vec{\theta},\vec{P};\,l;\,\theta,\,P) = \langle \chi(\vec{\theta},\vec{P};\,l;\,\theta,\,P)\vert\hat{O}\vert \chi(\vec{\theta},\vec{P};\,l;\,\theta,\,P)\rangle.
		\label{eq:O_new_exp}
	\end{equation}
	the expectation value of any operator $\hat{O}_i$ in the new path, that is  $O_i(\vec{\theta},\vec{P};l;\theta,P)$ varies sinusoidally with $2\theta$ \cite{nakanishi2020sequential,ostaszewski2021structure}:
	\begin{equation}
		O_i(\theta)=\left(a_i+ b_i\cos(2\theta) + c_i\sin(2\theta)\right),
		\label{eq:operator_three_point}
	\end{equation}
	This is derived in the Appendix.  The coefficients $a_i,b_i,c_i$ depend on $\vec{\theta},\vec{P}$, the location $l$ , the Pauli $P$, and the initial state $\ket{\chi}$. 
	The corresponding cost function is:
	\begin{equation}
		C(\vec{\theta},\vec{P};l;\theta,P):= \sum_{i=1}^N w_i\left(O_i(\vec{\theta},\vec{P};l;\theta,P) - \tau_i\right)^2.
		\label{eq:cost_derived_path}
	\end{equation}
	For an $n-$qubit system, there are $4^n$ distinct Pauli strings $\mathcal{P}_{all}$, and evaluating all of above expectation values requires exponential number of calls to the Oracle. To mitigate this, we restrict our attention to a much smaller set $\mathcal{P}_{pool}\subset \mathcal{P}_{all}$ of Pauli strings. 
	\subsection{QUEST Implementations}
	A psuedo-code for the QUEST is given below
	\begin{algorithmic}[1]
		\Statex \textbf{QUEST Algorithm}
		\Require Pauli pool $\mathcal{P}$, operators and corresponding targets $\{(O_i,\tau_i)\}$, initial state $|\chi\rangle$
		\State Initialize $|\psi\rangle \leftarrow |\chi\rangle$, $\vec{P} \leftarrow ()$, $\vec{\theta} \leftarrow ()$, and cost $C$.
		\Repeat
		\State Find $(P^\ast,l^\ast,\theta^\ast)$ and add $P^\star$ oriented at $\theta^\star$  at the location $l^\star$. Update $(\vec{P},\vec{\theta})$ and cost C.
		\State Update $\vec{\theta}$ using classical optimization, and update the cost $C$
		\Until{convergence of $C$}
		\State \Return $(\vec{\theta},\vec{P})$
	\end{algorithmic}

	Every iteration of a QUEST implementation involves two phases:
	\begin{enumerate}
		\item \textbf{Insertion Phase }: Insert a Pauli $P^\star$ at an angle $\theta^\star$ in the location $l^\star$ of the Pauli path.
		\item \textbf{Optimization Phase}: Do classical optimization of the angles $\vec{\theta}$. 
	\end{enumerate}
	In the Insertion Phase, identifying the best Pauli and location pair $(P^\star, l^\star)$ is very crucial. In the QUEST framework, we identify two different methods of selecting Pauli $P^\star$ and two different methods of selecting location $l^\star$, as shown in Table $\ref{tab:four_implementations}$. In the first two implementations (QUEST-tG and QUEST-tE), the new Pauli string is always inserted at the terminal, that is, the end, of the Pauli Path. The last two implementations (QUEST-bG and QUEST-bE) involve inserting the new Pauli at the best possible location in the Pauli path. Similarly, there are two  ways of selecting $P^\star$, the first method estimates the gradient of the cost function with respect to $\theta$ and uses it to select $P^\star$, while the second method evaluates the cost function at several angles for each Pauli string and selects the one that analytically minimizes the cost function.
	\begin{table}[h]
		\caption{Four algorithms implemented in the QUEST framework}
		\label{tab:four_implementations}
		\begin{tabular}{|c|c|c|}
			\hline
			Selection of location $l^\star$& Selection of Pauli $P^\star$ & Name  \\
			\hline
			Terminal& Gradient &QUEST-tG  \\
			\hline
			Terminal& Exact &QUEST-tE  \\
			\hline
			Best& Gradient &QUEST-bG  \\
			\hline
			Best& Exact &QUEST-bE  \\
			\hline
		\end{tabular}
	\end{table} 
	\par
	For an $n-$qubit system, there are $4^n$ distinct Pauli strings $\mathcal{P}$. Evaluating the expectation values for all the Pauli strings requires exponential number of calls to the Oracle. To mitigate this, we restrict our attention to a much smaller set $\mathcal{P}_{pool}$.
	\begin{equation}
		\mathcal{P}_m = \left\{ P\in\mathcal{P},\textbf{wt}(P)\le m\right\}
	\end{equation}
	\subsubsection{QUEST tE and bE implementations}
	
	The key property that makes analytic parameter determination possible is that  following: \par
	The cost functions of Eq. (\ref{eq:cost_derived_path}), in case where rest of the parameters are fixed, therefore becomes:
	\begin{equation}
		C(\theta) = \sum_{i=1}^N w_i\left(O_i(\theta) - \tau_i\right)^2.
		\label{eq:cost_theta}
	\end{equation}
	Here, $C(\theta)$ is $C(\vec{\theta},\vec{P};l;\theta,P)$, seen only as a function of $\theta$. This cost function $C(\theta)$ depends on $\theta$ as:
	\begin{equation}
		C(\theta) = \alpha + \beta \cos(2\theta) + \gamma \sin(2\theta)
		+ \delta \cos(4\theta) + \epsilon \sin(4\theta),
		\label{eq:C_expansion}
	\end{equation}
	where $\alpha,\beta,\gamma,\delta,\epsilon$ are scalar coefficients determined by $\{a_i,b_i,c_i,\tau_i\}$.  This five-parameter form is a direct consequence of squaring and summing the three-parameter expectation values of Eq. (\ref{eq:operator_three_point}), where the cross terms $\cos^2(2\theta)$, $\cos^2(2\theta)$, and $\cos(2\theta)\sin(2\theta)$ introduce the $4\theta$ harmonics. 
	
	Since $C(\theta)$ is a trigonometric polynomial  with five unknown coefficients, it is fully determined by evaluating $C(\theta)$ 
	at any five angles for which the resulting linear system is nonsingular. One such set is the symmetric one:
	\begin{equation}
		\Theta = \left\{\frac{k\pi}{5} \;:\; k = -2, -1, 0, 1, 2\right\},
		\label{eq:optimal_angles}
	\end{equation}
	From the five evaluations $\{C(k\pi/5)\}_{k=-2}^{2}$, 
	the coefficients are recovered via:
	\begin{equation}
		\begin{pmatrix} \alpha \\ \beta \\ \gamma \\ \delta \\ \epsilon \end{pmatrix} 
		= M^{-1} 
		\begin{pmatrix} C(-2\pi/5) \\ C(-\pi/5) \\ C(0) \\ C(\pi/5) \\ C(2\pi/5) 
		\end{pmatrix},
		\label{eq:coeff_recovery}
	\end{equation}
	where the matrix $M$ has rows indexed by $\theta \in \Theta$ and columns 
	corresponding to $[1,\ \cos 2\theta,\ \sin 2\theta,\ \cos 4\theta,\ 
	\sin 4\theta]$:
	\begin{equation}
		M 
		= \begin{pmatrix}
			1 & c_2 & -s_2 & c_4 & -s_4 \\
			1 & c_1 & -s_1 & c_2 & -s_2 \\
			1 & 1   &  0   & 1   &  0   \\
			1 & c_1 &  s_1 & c_2 &  s_2 \\
			1 & c_2 &  s_2 & c_4 &  s_4
		\end{pmatrix},
		\label{eq:M_matrix}
	\end{equation}
	where $c_j \equiv \cos(2j\pi/5)$ and $s_j \equiv \sin(2j\pi/5)$ for $j \in \{1, 2, 4\}$.
	The angle set $\Theta$ of Eq. (\ref{eq:optimal_angles}) is optimal in the sense that the condition number of the corresponding $M$ of Eq. (\ref{eq:M_matrix}), $\kappa(M)=\sqrt{2}$ is the minimum possible, ensuring minimal amplification of shot noise from oracle evaluations to the recovered  coefficients.\par
	Once the coefficients $\{\alpha,\beta,\gamma,\delta,\epsilon\}$ are determined, $C(\theta)$ of Eq. (\ref{eq:C_expansion}) is known analytically. The global 
	minimum $\theta^\star = \arg\min_\theta C(\theta)$ is obtained by evaluating 
	$C(\theta)$ on a dense uniform grid over $[-\pi, \pi]$ and selecting the minimizer. 
	Since $C(\theta)$ is a low-frequency trigonometric polynomial with known 
	coefficients, a grid of $N_g$ points resolves the global minimum to within 
	$\pi/N_g$ radians, with no risk of missing the global minimum due to aliasing,  as the highest frequency present is $4\theta$. 
	
	\subsubsection{QUEST tG and bG Implementations}
	We select the best Pauli $P^\star$ and the best location $l^\star$ as those which yield the largest gradient of cost in the Pauli Path $\vec{\theta},\vec{P}$.  For this, we define a quantity called $g(l;P)$ as the gradient of the cost $C(\vec{\theta},\vec{P};l;\theta,P)$ evaluated at $\theta=0$:
	\begin{equation}
		g(l;P)=\frac{d}{d\theta}C(\vec{\theta},\vec{P};l;\theta,P)\vert_{\theta=0}
	\end{equation}
	Here $g(l;P)$ also depends on the current state $\vec{\theta}$ and Pauli string $P$, but it is not explictly written for notational simplicity. The derivative of the cost is given by:
	\begin{equation}
		g(l;P)=2\sum_{i=1}^N w_i\left(O_i(\vec{\theta},\vec{P}) - \tau_i\right)\frac{d}{d\theta}O_i(l;\theta,P)\vert_{\theta=0},
		\label{eq:gradient_formula}
	\end{equation}
	where $O_i(\vec{\theta},\vec{P})$ is the expectation value of $\hat{O}_i$ in the current state, see Eq. (\ref{eq:operator_expectation_value}). 
	The derivative appearing in the RHS can be computed by noting that $O_i(l;\theta,P)$ varies sinusoidally with $\theta$, as in Eq. (\ref{eq:operator_three_point}). Therefore the derivate can obtained analytically by evaluating $O_i(l;\theta,P)$ at two values of $\theta$. 
	\begin{equation}
		\frac{d}{d\theta}O_i(l;\theta,P)\vert_{\theta=0}=O_i\big(l;\frac{\pi}{4},P\big)-O_i\big(l;-\frac{\pi}{4},P\big).
	\end{equation}
	In the QUEST-bG implementation, the $l^\star$ and $P^\star$ are chosen as $l^\star,P^\star=argmax(\abs{g(l;P)})$. 
	The selected Pauli $P^\star$ oriented at the angle $0$ is mounted at the location $l^\star$.  In the QUEST-tG implementation, the $l^\star$ is always the terminal position $t$ of the Pauli path and the $P^\star$  is selected as $P^\star=argmax(\abs{g(t;P)})$.

	\subsubsection{Classical Optimization}
	\label{sec:classical_opt}
	
	In the Optimization Phase, all the  accumulated rotation angles in the current Pauli Path, $\vec{\theta}$, are jointly optimized. In this phase the goal is to minimize the full multi-constraint cost function $C(\vec{\theta}, \vec{P})$ with respect to the complete angle vector $\vec{\theta}$, while keeping the Pauli string sequence $\vec{P}$ fixed. This classical refinement of angles is implemented across all the four QUEST implementations. This step performs two essential roles: (i) it compensates for the inherent sub-optimality single-angle insertion criterion, and (ii) it enables previously inserted rotations to adapt cooperatively to the newly modified constraint landscape. Gradients required for this phase, if not already available, are evaluated analytically via the parameter-shift rule~\cite{wierichs2022general}. Between the gradient-based QUEST methods  and the exact-based QUEST methods, it is expected that the classical optimizers perform better in the latter case, as the newly inserted Pauli $P^\star$ is already oriented at the best angle $\theta^\star$, whereas in the former case the newly inserted Pauli $P^\star$ is oriented at $0$ angle.

	The insertion phase and classical optimization phase together constitute one iteration of the QUEST algortihm. An iteration of QUEST typically leads to an increase in the Pauli path length by one. However, if an angle $\theta_l \approx 0$, one could remove the unitary $e^{i\theta_lP_l}$ from the path, leading to reduction in the Pauli path length. 
	\subsubsection{Stopping Condition of QUEST Implementations}
	QUEST terminates at the iteration where the cost function becomes close to $0$. The terminal state at this stage is the desired state that satisfies the constraints of Eq. (\ref{eq:target_equation}). It is, however, possible that the cost remains finite (non-zero), but QUEST finds that insertion of any Pauli string $P\in\mathcal{P}_{pool}$ at any location in the Pauli path only increases the cost further. We say a pool $\mathcal{P}_{pool}$ is saturated if it is not  possible to reduce the cost any further by using any Pauli $P$ from that pool. In that case, one can augment the pool with Pauli strings of larger weights. 
	
	It can also happen that no Pauli string of any weight reduces the cost any further. This is more likely to happen in QUEST-tG implementation, owing to the reasons discussed earlier. In such a scenario, one could repeat the algorithm starting from a different initial state. 
	
	If none of the tweaks help in reducing the cost beyond a certain value, it indicates that the given set of constraints are inconsistent, and there is no single pure state $\ket{\phi}$ that satisfies all the constraints of Eq. (\ref{eq:target_equation}). Such QUEST runs are still important.
	The terminal state corresponding to the least possible cost indicates the best state in the least-square sense for a given set of constraints. 
	\subsection{Comparision with Existing Methods}
	The QUEST-tG implementation is the closest in spirit to the ADAPT-VQE algorithm employed for Hamiltonian minimization. QUEST-tG therefore inherits the strengths and weakness of ADAPT-VQE. However, the requirement of multiple constraints satisfaction adds additional complexity to QUEST-tG, not present in ADAPT-VQE. 
	In Hamiltonian minimization, the cost is $C = \langle\psi(\vec{\theta})|H| 
	\psi(\vec{\theta})\rangle$ directly, and the gradient reduces to 
	$dC/d\theta_P|_{\theta_P=0} = i\langle\psi|[H,P]|\psi\rangle$ directly. The pool element for which $|\langle\psi|[H,P]|\psi\rangle|$ is largest is selected. This expectation value of the commutator vanishes  only when $|\psi\rangle$ is an eigenstate of $H$ which is precisely the termination condition. One can therefore say that the gradient magnitude and progress toward the ground state are  unambiguously coupled in ADAPT-VQE. In the present  multi-constraint setting, the quadratic cost of  Eq.~\eqref{eq:vqe_cost_target} is necessary to reduce $N$ simultaneous constraints of Eq. (\ref{eq:target_equation}) to a single scalar objective, and the gradient Eq.~\eqref{eq:gradient_formula} acquires the residuals $w_i(\langle\psi(\vec{\theta})|O_i|\psi(\vec{\theta})\rangle - \tau_i)$ as  signed weights. These residuals introduce failure modes absent from Hamiltonian 
	minimization: they can cancel across observables in Eq.~\eqref{eq:gradient_formula},  driving the total gradient to zero while the state satisfies none of the  constraints; the pool element with the largest gradient magnitude need not be  the one that most reduces $C$ at the optimal finite angle, since the gradient at $\theta_P = 0$ reflects only the local slope of a trigonometric landscape  whose global minimum may lie far from the origin, and as the circuit grows,  the residuals evolve at every iteration, creating a non-stationary selection landscape in which a previously effective pool element becomes ineffective due  to redistribution across constraints.
	
	It is also to be noted that unlike the energy-minimization problem, where the pursuit is for a specific state (the ground state of the Hamiltonian) upto a global phase, the problem of expectation value targetting generally admits multiple pure states as solutions. Many states can satisfy the given set of constraint (Eq (\ref{eq:target_equation}) ), and the algorithm can land on any of these. Therefore it is likely that the gradient-based methods can perform better in this class of problems than in the energy minimization problem. 
	\subsection{Role of the weights.}
	The positive weights $\{w_i\}$ appearing in Eq. (\ref{eq:cost_function}) need not sum up to 1, but owing to the invarance of cost function with respect to scaling of weights $\textbf{w}\rightarrow \lambda\textbf{w}$, for any $\lambda>0$, we choose the weights $\textbf{w}$ such that $mean(\textbf{w})=1$, without any loss of generality. 
	
	One problem context where different weigths are assigned to different operators is when the operators have different spectral widths. 
	Let $\lambda_{\max}(\hat{O})$ and $\lambda_{\min}(\hat{O})$ be the largest and smallest eigenvalues of the operator $\hat{O}$. Define the spectral width as $\Delta(\hat{O}) = \lambda_{\max}(\hat{O}) -\lambda_{\min}(\hat{O})$. A cost-function with uniform weight will only drive the expectation value of the operator $\hat{O}_j$ with the largest spectral width $\Delta(\hat{O}_j)$ to its target value $\tau_j$, while the other operators stay away from their targets, as there is no cost incentive to bring them to their targets.  To counter this, the weights can be chosen as $w_j = 1/\Delta_j^2$. This rescales each
	residual by the maximum value it could take, so that
	$(\langle O_i \rangle_\theta - t_i)^2 / \Delta_i^2 \in [0,1]$ regardless of the operator's spectral widths. 
	
	The weights can also be assigned to reflect the relative importance of each operator appearing inside the cost function.
	For example, higher weights can be assigned to prioritize non-negotiable observables such as symmetries and conserved quantities, while other observables can be assigned smaller weights.
	
	Another way to assign weights is to take into account the statistical nature of the target values. This situation arises when the target values are not known exactly but are known with some uncertainty. 
	When each target $\tau_i$ is specified with an associated uncertainty $\sigma_i$, the weights can be directly selected as $w_i \;=\; \frac{1}{\sigma_i^2}$.
	This way, targets specified with higher confidence (smaller $\sigma_i$) carry larger weight in the cost, while loosely specified targets (larger $\sigma_i$) contribute less and are permitted to deviate further from the achieved expectation value at convergence.

	\subsection{Oracle Evaluations in QUEST Implementations}
	
	We now consider the number of Oracle calls in the exact Insertion QUEST implementations. 
	Consider the $t^{th}$ iteration. The length of current Pauli path is $(t-1)$. 
	\subsubsection{QUEST-bE and QUEST-tE Implementations}
	Consider first the QUEST-bE implementation. In the insertion phase, for identifying $P^\star$ and $l^\star$ the cost $C(\vec{\theta},\vec{P};\,l;\,\theta,P)$ is evaluated at five angles $\theta\in \Theta$, for every Pauli string $P\in\mathcal{P}_n$, and at every insertion position $l\in\{0,1,\cdots,t-1\}$. However, the cost $C(\vec{\theta},\vec{P};\,l;\,\theta,P)$ at $\theta=0$ is independent of $P$ and $P$ and is equal to the current cost. That is $C(\vec{\theta},\vec{P};\,l;\,\theta=0,P)=C(\vec{\theta},\vec{P})$. Therefore the cost is evaluated for only four angles. The total number of Oracle calls in this phase at the$t^{th}$ iteration is $4\times N\times t\times\abs{\mathcal{P}_{pool}}$.\\
	In case of QUEST-tE, since we check for insertion only at the terminal position, the Oracle calls is $4\times N\times \abs{\mathcal{P}_{pool}}$
	\subsubsection{QUEST-bG and QUEST-tG Implementations}
	We now consider the number of Oracle calls in the $t^{th}$ iteration in the gradient based QUEST implementations. Consider first the QUEST-bG implementation. Here, in the insertion phase, $P^\star$ and $l^\star$ are identified through $l^\star,P^\star=argmax(\abs{g(l;P)})$. Computing $g(l;P)$  requires evaluating $O_i(l;\theta,P)$ at two angles $\theta=\frac{\pi}{4}$ and $\theta=-\frac{\pi}{4}$, for all the constraints $i=1,\cdots,N$, requring $2N$ oracle calls. The quantity $g(l;P)$ is evaluated for all insertion locations $l=0,\cdots,t-1$ and for all Pauli strings $P\in\mathcal{P}_{pool}$. Therefore the total number of Oracle calls is $2\times N\times t\times \abs{\mathcal{P}_{pool}}$
	
	In case of QUEST-tG, since we check for insertion only at the terminal position, the total number of Oracle calls per iteration is $2\times N\times \abs{\mathcal{P}_{pool}}$.

	\section{Results}
	QUEST can also be used to solve conventional problems of quantum state engineering, by using it in a single operator single target mode. For instance, it is possible to synthasize a pure $n$-qubit state $\ket{\phi}$ directly using the expectation-value targeting formalism introduced here, by choosing the operator $\hat{O}_1$ as the corresponding density matrix $\rho=\ket{\phi}\bra{\phi}$, with the target value $\tau_1$ equal to 1, and weight $w_1=1$. Likewise QUEST can be used for identifying the ground state of the Pauli string Hamiltonians $\hat{H}$, just by changing the cost function to $C(\vec{\theta}, \vec{P})=\langle \chi(\vec{\theta},\vec{P})|\hat{H}|\chi(\vec{\theta},\vec{P})\rangle$.
	
	In this Section, we give numerical illustration of genuiene multiple operator expectation-value targeting with all the four implementations of QUEST, and compare there performace. We employ the L-BFGS quasi-Newton optimizer~\cite{liu1989limited} for classical optimization of the cost function $C(\vec{\theta}, \vec{P})$.
	
	We define the root mean square (RMS) cost per constraint as $C_{RMS} = \sqrt{\frac{1}{N}\sum_{i=1}^N w_i(\langle O_i \rangle-\tau_i)^2}$. The runs are terminated when either this $C_{RMS}$ is reduced below a pre-defined threshold $\epsilon$, or the maximum number of iterations $I_{max}$ is reached.
	
	\subsection{State Synthesis from Pauli Expectation Values}
	In this first illustration, we select as the operators $\{\hat{O}_i\}$ to be the $N $ random Pauli strings, and run QUEST to obtain the set targets. 
	
	To ensure that the constraints are consistent with each other, we first prepare a random state $\ket{\psi_0}$ and then compute the expectation values of the operators with respect to this state. These expectation values are then set as the target values for the algorithm. This ensures that that there is atleast one state that yeilds the desired expectation values. 
	
	We study this problem in two different scenarios: (i) noise-less case and (ii) the case with simulator noise. 
	
	In the former case, it is assumed that all the targets are known with perfect accuracy. In the latter case, targets are assumed to be computed through a finite number of measurements of the respective Pauli strings on the reference state $\ket{\psi_{ref}}$. In that case, the target is given by the mean value $\tau_i$ of the measurement outcomes of $P_i$ on the reference state $\ket{\psi_{ref}}$. 
	
	The standard deviation this estimate is given by 
	\begin{equation}
		\sigma_i \;=\; \sqrt{\frac{1 - \tau_i^2}{M_i}},
		\label{eq:std_noisy_Pauli_String}
	\end{equation}
	where $M_i$ is the number of measurements performed for estimating the expectation value of the $i^{th}$ Pauli string. 
	
	In QUEST implementation of this scenario, we assign the weights as $w_i=\frac{1}{\sigma_i^2}$ for all $i=1,\dots,N$. Since there is a freedom of the overall scaling of the weights, we set scale the weights such that $\sum_{i=1}^N w_i = N$. 
	
	We now study the case of a $10$-qubit system for which $100$ random Pauli strings are chosen as the operators, with the target set as those obtained by a Haar random pure state on $2^{10}$-dimensional Hilbert space. 
	All four implmentations of QUEST were run from the same initial state $\ket{0}^{\otimes n}$, with the Pauli pool being the set of all weight-$2$ Pauli strings $\mathcal{P}_2$. The RMS cost per constraint was set at $\epsilon=10^{-3}$.
	
	\subsubsection{Noise-less case}
	
\begin{figure}[h]
		\includegraphics[width=\columnwidth]{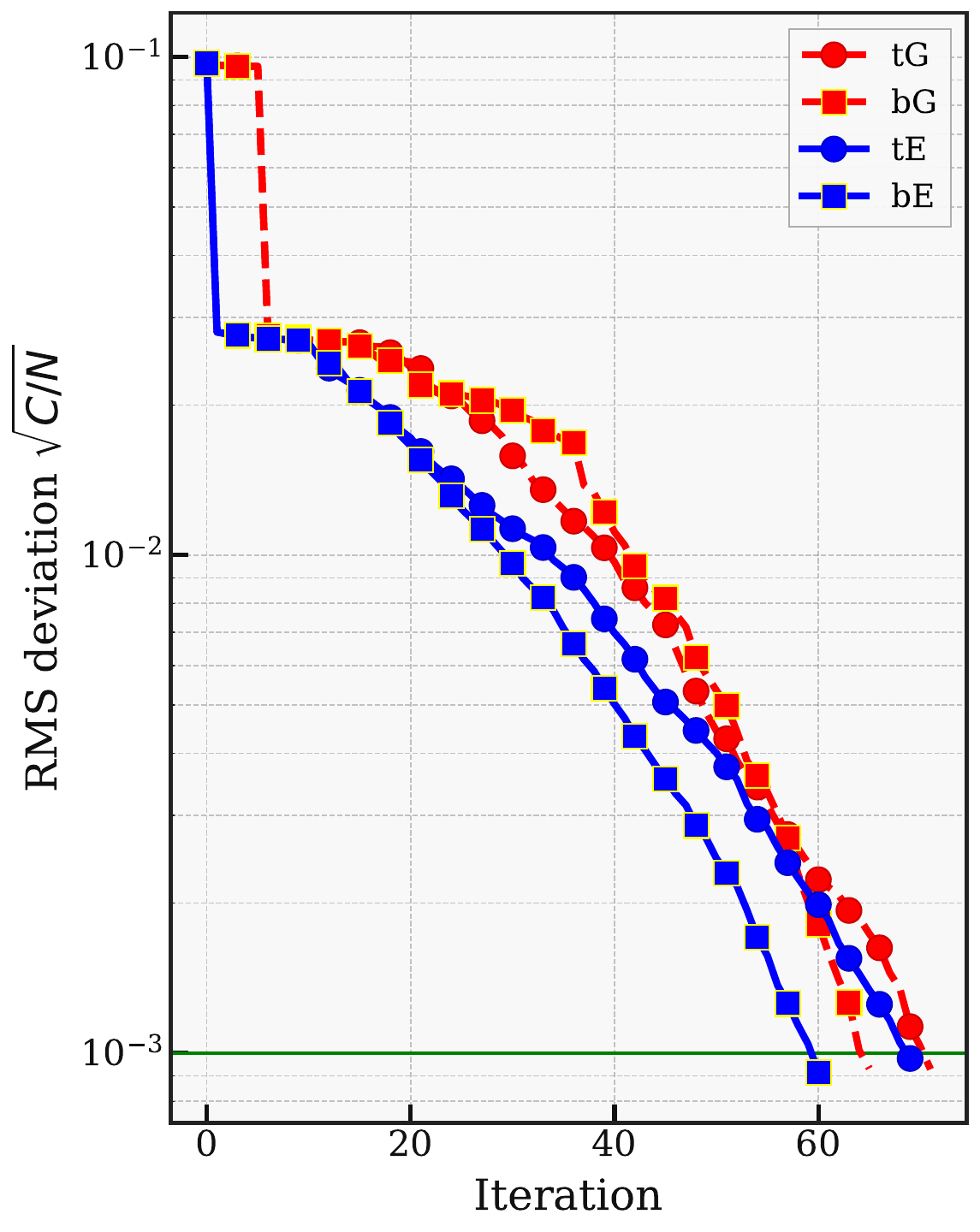}
		\caption{Evolution of RMS cost per constraint for the four implementations of QUEST for the noise-less random Pauli string expectation value targeting of $n=10$ qubit, with $100$ constraints on the expectation values of random Pauli strings. } 
		\label{fig:noiselss_RMS}
	\end{figure}

	The RMS cost per constraint at every iteration in the four versions of QUEST are shown in Fig. (\ref{fig:noiselss_RMS}).
	
	\begin{figure}[h]
		\includegraphics[width=\columnwidth]{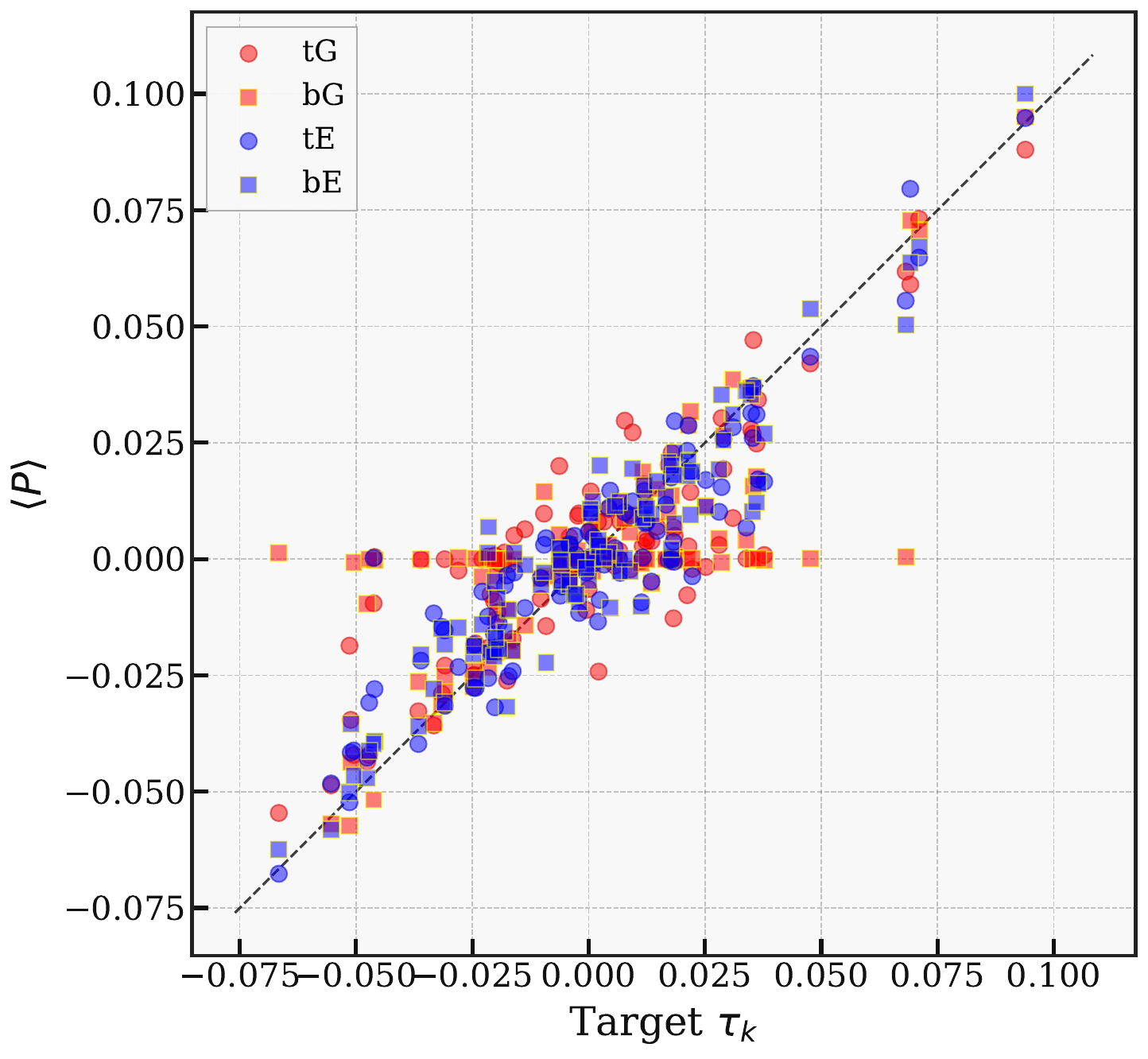}
		\caption{Scatter plot of set targets vs prepared expectation values for the four implementations of QUEST, at the end of 30 iterations. The diagonal line $y=x$ is shown for reference. } 
		\label{fig:noiseless_scatter}
	\end{figure}

	The QUEST-bG implementation reaches the desired RMS cost per constraint $\epsilon = 10^{-3}$ in about $t=60$ iterations, the quickest of the four. While the QUEST-tG implementation takes about $10$ iterations more to reach the same $\epsilon$, being the slowest of the four. The terminal states of all the four cases are such that the expectation values of each of the $100$ Pauli strings match their targets up to an error of $\pm 10^{-3}$ on average. It is however instructive to compare the actual and targeted expectation values of the four implementations after the same number of iterations. In Fig. (\ref{fig:noiseless_scatter}), we present a scatter plot of the $100$ targets and the corresponding expectation values obtained in the terminal states of the four implementations of QUEST. It is evident from the scatter plot that the gradient-free implementations, QUEST-tE and QUEST-bE (blue circles and blue squares), are significantly more accurate in matching the target expectation values compared to the gradient-based implementations, QUEST-tG and QUEST-bG (red circles and red squares).
	\subsubsection{Noisy case}
\begin{figure}[h]
		\includegraphics[width=\columnwidth]{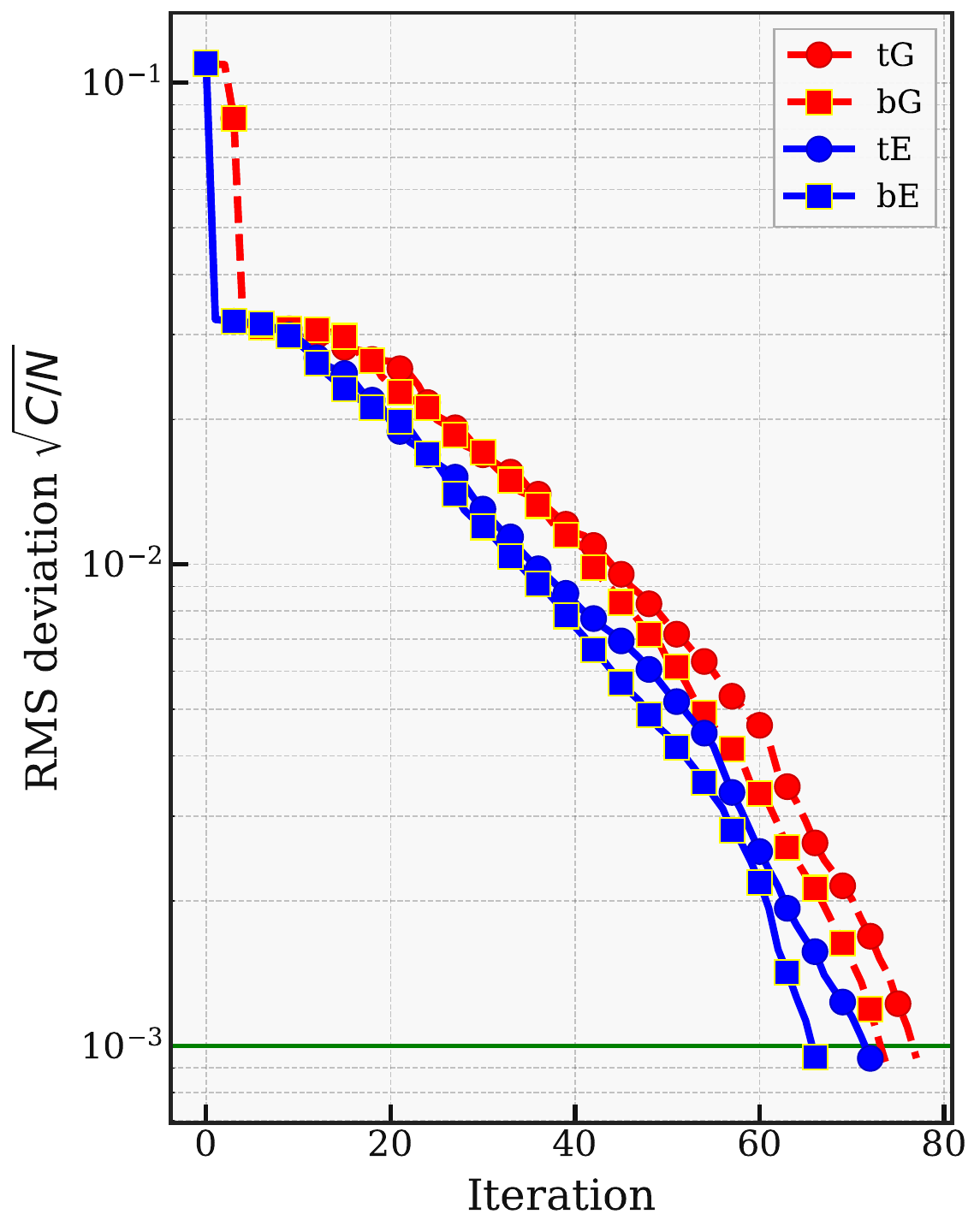}
		\caption{Evolution of RMS cost per constraint for the four implementations of QUEST for the noisy random Pauli string expectation value targeting of $n=10$ qubit, with $100$ constraints on the expectation values of random Pauli strings. } 
		\label{fig:noisy_RMS}
	\end{figure}

\begin{figure}[h]
		\includegraphics[width=\columnwidth]{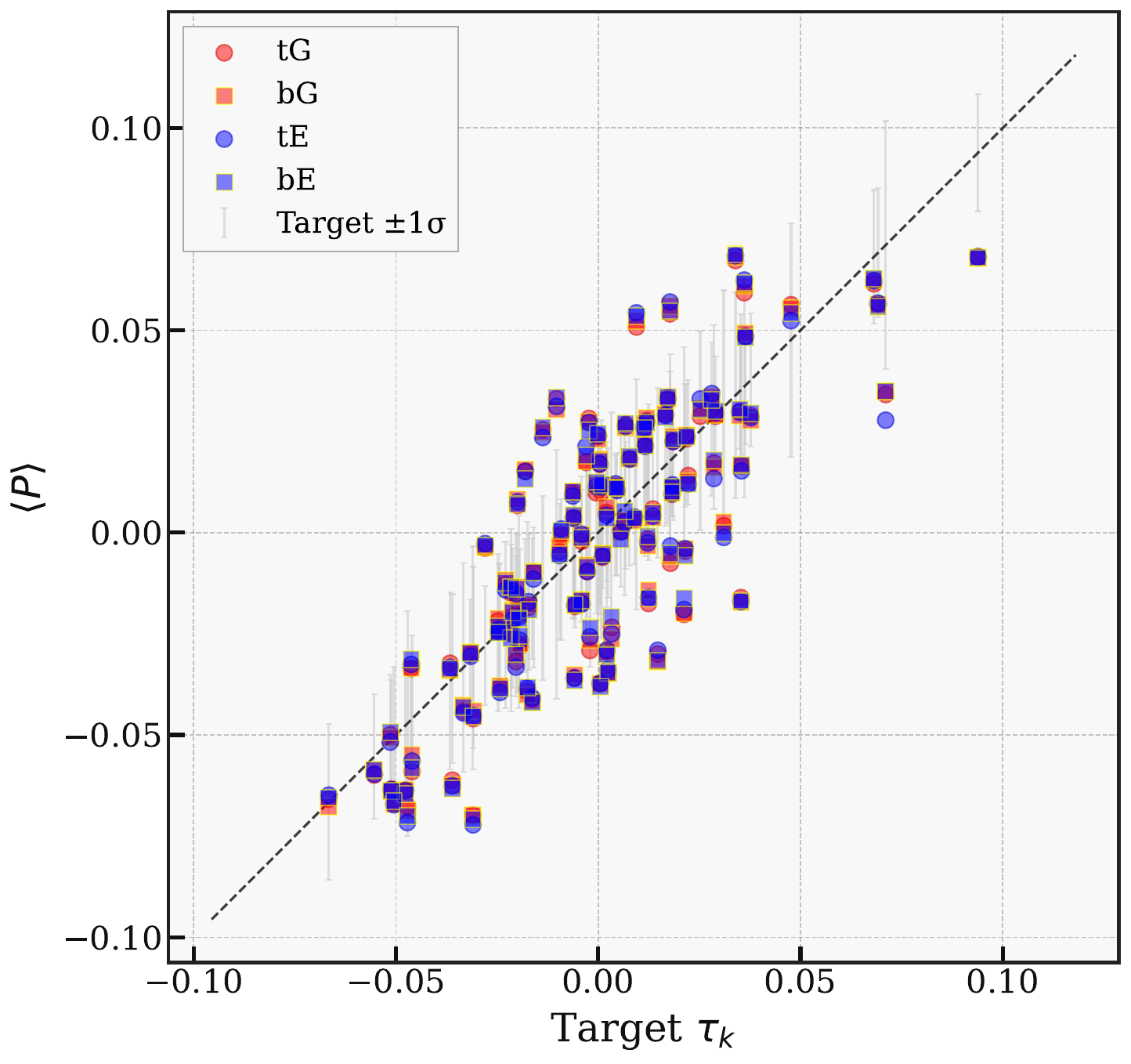}
		\caption{Scatter plot of set targets vs prepared expectation values for the four implementations of QUEST, computed in the terminal state of the four implementations. The diagonal line $y=x$ is shown for reference. The grey vertical lines correspond to the $\pm \sigma_i$ bounds around the targets $\tau_i$. } 
		\label{fig:noisy_scatter}
	\end{figure}

	In this illustration, we take the same reference state as in the noise-less case, the same set of $N=100$ Pauli strings, and consequently the same targets $\tau_i$. However, these targets are now assumed to be noisy. To simulate the noisy targets $\tau_i$, we draw $M_i, i=1,\cdots,N$ randomly between $1000$ and $5000$. The corresponding standard deviation $\sigma_i$ are computed using Eq. (\ref{eq:std_noisy_Pauli_String}). 
	Figure (\ref{fig:noisy_RMS}) shows the evolution of RMS cost per constraint for the four implementations of QUEST in this noisy case. As in the noise-less case, the QUEST-bE implmentation reaches the desired RMS target per constraint $\epsilon = 10^{-3}$ the quickest of the four, although it takes a few iterations longer than in the noise-less case. Owing to the inherent noise in the target-values $\tau_k$, even at the target RMS per constraint of $\epsilon=10^{-3}$, the actual expectation values $\langle P_k \rangle$ deviate from the target values $\tau_k$. This is captured in the Fig. (\ref{fig:noisy_scatter}). Note that unlike Fig. (\ref{fig:noiseless_scatter}), which corresponds to the $30^{th}$ iteration, the scatter plot in Fig. (\ref{fig:noisy_scatter}) corresponds to the terminal states of the four implementations of QUEST.
	
	\begin{figure}[htbp]
		\includegraphics[width=1\columnwidth]{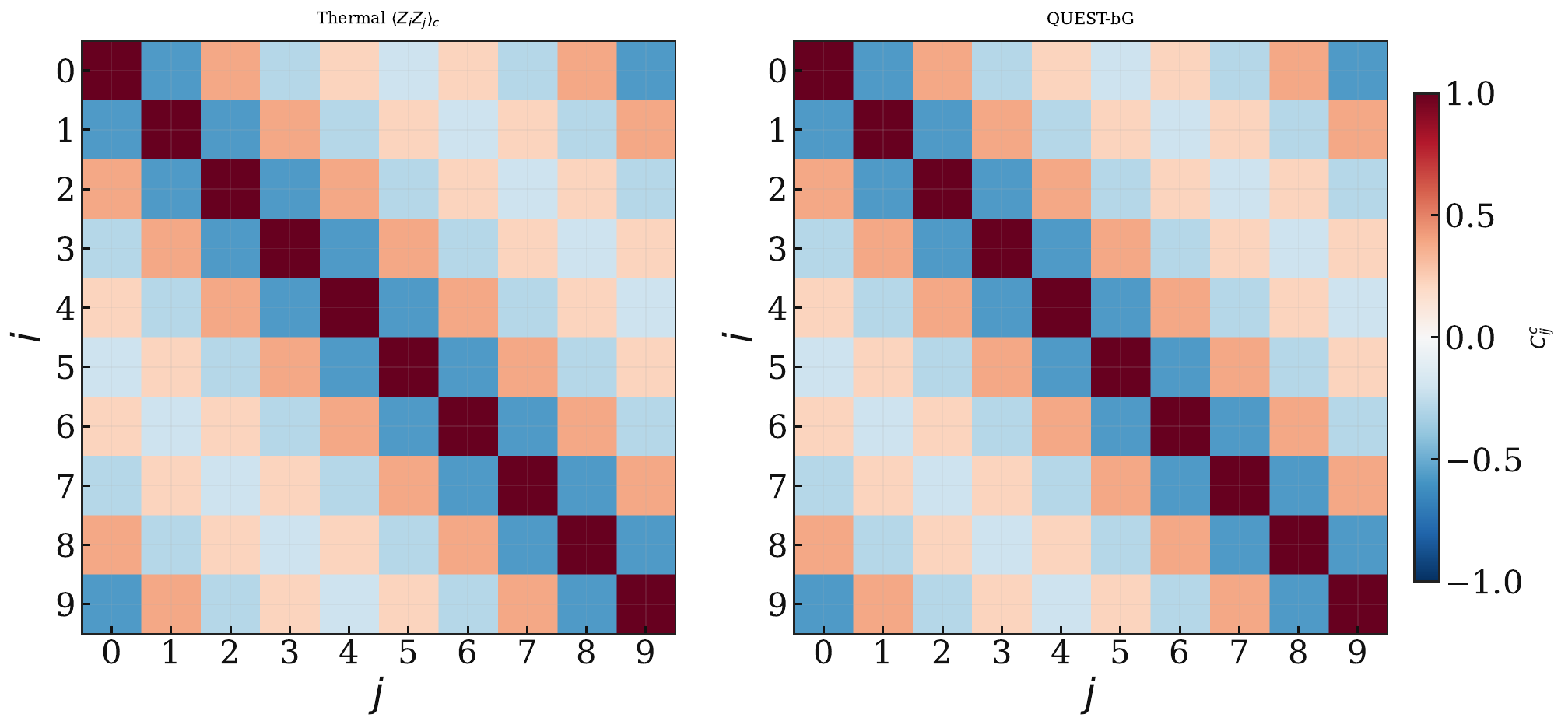}
		\caption{Demonstration of closest pure state to a thermal state at $\beta=1$, for the TFIM of Eq. (\ref{eq:H_TFIM}), with $J=-1$  and $h=1$ with $n=10$ qubits. Left panel depicts the correlation $C_\beta(i,j)$ defined in Eq. (\ref{eq:beta_correlation}) and the right panel depicts the correlation $C_\psi(i,j)$ defined in Eq. (\ref{eq:correlation_psi}), where $\ket{\psi}$ is the state synthesized by QUEST-bG at the terminal iteration. } 
		\label{fig:correlation_replica}
	\end{figure}
	\begin{figure}[htbp]
		\includegraphics[width=1\columnwidth]{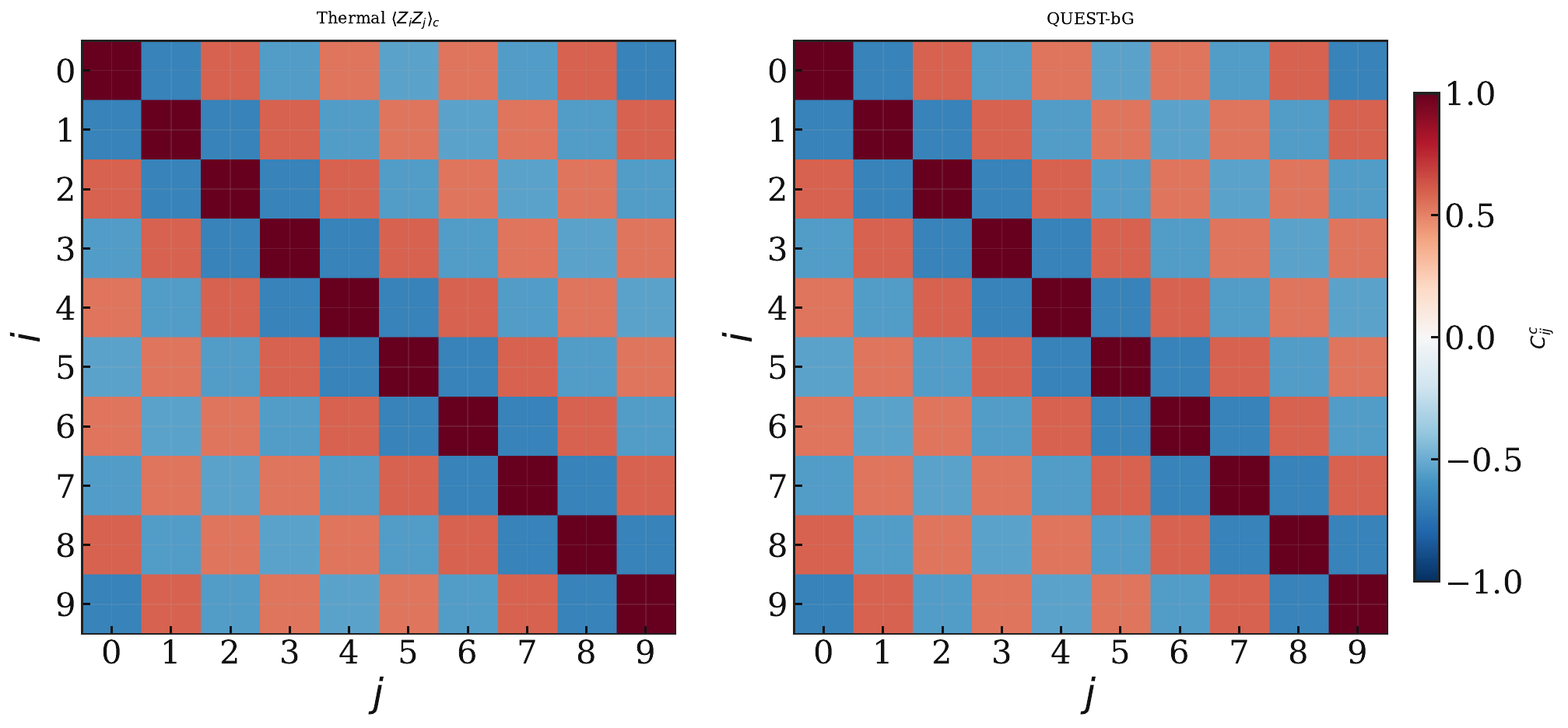}
		\caption{Demonstration of closest pure state to a thermal state at $\beta=3$, for the TFIM of Eq. (\ref{eq:H_TFIM}), with $J=-1$  and $h=1$ with $n=10$ qubits. Left panel depicts the correlation $C_\beta(i,j)$ defined in Eq. (\ref{eq:beta_correlation}) and the right panel depicts the correlation $C_\psi(i,j)$ defined in Eq. (\ref{eq:correlation_psi}), where $\ket{\psi}$ is the state synthesized by QUEST-bG at the terminal iteration. } 
		\label{fig:correlation_replica2}
	\end{figure}
	
	
	\subsection{Preparation of a Closest Pure State }
	\label{sec:pseudo_thermal}
	Here we address the following problem:
	\begin{quote}
		\emph{Consider a mixed state $\rho$, and a set of observables ${O_i}$.  Compute their expectation values  in this mixed state $\rho$, i.e., $t_i = \langle O_i \rangle_{\rho}$. Then, try to find a pure state $\ket{\psi}$ such that}
		\begin{equation}
			\langle \psi | O_i | \psi \rangle \approx t_i \quad \forall i.
		\end{equation}
	\end{quote}
	We call such a state $\ket{\psi}$ as the \textit{closest pure state} to the mixed state $\rho$. The problem of finding the closest pure state to a given mixed state is a problem that is directly within the scope of expectation-value targeting. \par
	As the first illustration of this idea, we generate what we call pseudo-thermal states. A mixed thermal state at inverse temperature $\beta$, with respect to a given Hamiltonain $\hat{H}$ is described by the Gibbs density matrix  $\rho_\beta(\hat{H})$ as:
	\begin{equation}
		\rho_\beta(\hat{H}) = \sum_k p_k \ket{E_k}\bra{E_k},
		\qquad
		p_k = \frac{e^{-\beta E_k}}{\displaystyle\sum_j e^{-\beta E_j}},
		\label{eq:gibbs}
	\end{equation}
	where $\{E_k, \ket{E_k}\}$ are the eigenpairs of the Hamiltonian $\hat{H}$, and $p_k$ is the Boltzmann weight of the $k$-th eigenstate.
	The expectation value of an operator $\hat{O}$ at the inverse temperature $\beta$, with respect to the thermal state corresponding to the Hamiltonian $\hat{H}$ is given by
	\begin{equation}
		\langle \hat{O} \rangle_\beta
		= \operatorname{Tr}[\rho_\beta(\hat{H}) \hat{O}]=\sum_{k} p_k \langle E_k |\hat{O}|E_k\rangle,
	\end{equation}
	We intend to  engineer a \emph{pseudo-thermal pure state} $\ket{\psi}$ which a pure quantum state such that 
	\begin{equation}
		\langle \psi | \hat{O}_i | \psi \rangle = \langle \hat{O}_i \rangle_\beta, \quad \text{for } i=1,\ldots,N.
	\end{equation}
	This setup is closely related to the generalized Gibbs ensemble (GGE) framework \cite{jaynes1957information,rigol2007relaxation,vidmar2016generalized}, in which an equilibrium state is characterized by the expectation values of a set of conserved or slowly-varying observables, and reconstructed via maximum entropy. Whereas the GGE yields a mixed state consistent with the specified expectation values, the present construction seeks a pure state producing the same expectation values on the chosen operator set. Such a pure state need not exist for arbitrary choices of the reference Hamiltonian, inverse temperature, and operator set; the numerical feasibility observed below is empirical and does not establish general existence. \par
	Finding the psuedo-thermal state is directly the expectation-value targeting problem Eq. (\ref{eq:target_equation}), with the target values being $\tau_i = \langle \hat{O}_i \rangle_\beta$. 
	As an illustration of pseudo-thermal state generation, consider the Hamiltonian of the transverse field Ising model:
	\begin{equation}
		\hat{H}_{\mathrm{TFIM}} = -J \sum_{i=0}^{n-1} Z_i Z_{i+1 \bmod n}
		- h \sum_{i=0}^{n-1} X_i,
		\label{eq:H_TFIM}
	\end{equation}
	where $J$ is the Ising coupling, $h$ is the transverse-field strength, and $Z_i$, $X_i$ denote the Pauli Pauli strings where only one Pauli operator $Z$ or $X$ is acting on qubit $i$ and rest are all identities.\\
	
	Consider the real quantity $C_\beta(i,j)$ as the specific correlation between the qubit $i$ and qubit $j$ defined as:
	\begin{equation}
		C_\beta(i,j) = \langle \hat{Z}_i\hat{Z}_j\rangle_{\beta} -\langle \hat{Z}_i\rangle_{\beta}\langle \hat{Z}_j\rangle_{\beta},
		\label{eq:beta_correlation}
	\end{equation}
	We now intend to synthesize a pure state $\ket{\psi}$ such that 
	\begin{equation}
		\begin{aligned}
			\langle \psi | \hat{Z}_i \hat{Z}_j | \psi \rangle & = \operatorname{Tr}[\rho_\beta(\hat{H}) \hat{Z}_i\hat{Z}_j], \text {for all } i,j \text { and }\\
			\langle \psi | \hat{Z}_i | \psi \rangle &= \operatorname{Tr}[\rho_\beta(\hat{H}) \hat{Z}_i], \text {for all } i,
		\end{aligned}
	\end{equation}
	so that the correlation function $C_\psi(i,j)$ defined as:
	\begin{equation}
		C_\psi(i,j) = \langle \psi|\hat{Z}_i\hat{Z}_j|\psi\rangle -\langle \psi|\hat{Z}_i|\psi\rangle\langle \psi|\hat{Z}_j|\psi\rangle.
		\label{eq:correlation_psi}
	\end{equation}
	is identical to $C_\beta(i,j)$ defined in Eq. (\ref{eq:beta_correlation}). 
	For an $n-$qubit system, there are $n$ single-qubit operators and $\frac{1}{2}n(n-1)$ two-qubit operators, leading to a total of $n+\frac{1}{2}n(n-1) = \frac{1}{2}n(n+1)$ constraints. 
	We have carried out the simulation of $n=10$ case with $J=-1$ and $h=1$, and at two inverse temperatures $\beta = 1.0$ and $\beta = 3.0$. The initial state was $\ket{+}^{\otimes 10}$. The two dimensional color-coded correlation matrix is shown in Fig. (\ref{fig:correlation_replica}) for $\beta = 1.0$ and Fig. (\ref{fig:correlation_replica2}) for $\beta = 3.0$. In these figures, the left panel corresponds to the target thermal state, and the right panel corresponds to QUEST-bG generated pure states. It is evident from this picture that QUEST-bG succeeds in capturing all the $55$ expectation values correctly, and therefore producing the correlation matrix close to that of the thermal state, at both the temperatures. 
	
	
	\subsection{Fermionic lattice Hamiltonian}
	Quantum chemistry Hamiltonian are written as \cite{mcardle2020quantum,bauer2020quantum}:
	\begin{equation}
		\hat{H} = \sum_{pq} h_{pq} \hat{a}^\dagger_p \hat{a}_q + \frac{1}{2} \sum_{pqrs} (pq|rs) \hat{a}^\dagger_p \hat{a}^\dagger_q \hat{a}_r \hat{a}_s,
		\label{eq:hamiltonian}
	\end{equation}
	where $\hat{a}^\dagger_p$ and $\hat{a}_p$ are the creation and annihilation operators of the fermion in the orbital $p$, and $h_{pq}$ and $(pq|rs)$ are the one- and two-electron integrals, respectively. The Hamiltonian in Eq. \eqref{eq:hamiltonian} provides a general second-quantized framework for interacting fermionic systems, including both molecular electronic structure and lattice models. Rather than working with chemistry-specific orbital bases and integral truncations, we consider the one-dimensional Hubbard model as a structured specialization of Eq. \eqref{eq:hamiltonian}, in which only nearest-neighbor hopping and on-site interactions are retained. This choice preserves the essential algebraic structure of fermionic Hamiltonians while providing a controlled testbed for benchmarking the four flavours of QUEST.
	\label{subsec:numerics_hubbard}
	\subsubsection{System and Jordan--Wigner mapping}
	\label{subsec:system}
	We consider a one-dimensional Hubbard chain on $M$ sites with open boundary conditions, at half filling ($N_e = M$ electrons).  In second quantization, its Hamiltonian is \cite{arovas2022hubbard,cade2020strategies,bespalova2025simulating}
	\begin{equation}
		\hat{H}
		= -t \sum_{p=0}^{M-2}
		\sum_{\sigma \in \{\uparrow, \downarrow\}}
		\bigl( \hat{a}^\dagger_{p, \sigma} \hat{a}_{p+1, \sigma}
		+ \text{H.c.} \bigr)
		+ U \sum_{p=0}^{M-1} \hat{n}_{p, \uparrow}\, \hat{n}_{p, \downarrow} ,
		\label{eq:hubbard}
	\end{equation}
	where $\hat{a}^\dagger_{p, \sigma}$ creates an electron of spin $\sigma$
	on site $p$, $\hat{n}_{p, \sigma} = \hat{a}^\dagger_{p, \sigma}
	\hat{a}_{p, \sigma}$ is the number operator, $t > 0$ is the nearest-neighbour hopping amplitude, and $U > 0$ is the on-site Coulomb repulsion. 
	Here $M$ is the number of spatial orbitals (lattice sites). \par
	
	Equation~\eqref{eq:hubbard} corresponds to a quantum-chemistry Hamiltonian with only two non-zero classes of integrals,
	\begin{equation}
		h_{p q}
		= \begin{cases}
			-t & |p - q| = 1 , \\
			0  & \text{otherwise},
		\end{cases}
		\qquad
		(p q | r s)
		= U \, \delta_{p r} \delta_{q r} \delta_{r s},
		\label{eq:integrals}
	\end{equation}
	with all other one- and two-body integrals vanishing.  \par
	We map the Hubbard Hamiltonian Eq.~\eqref{eq:hubbard} to a system of $n = 2 M$ qubits using the
	Jordan--Wigner (JW) transformation \cite{nielsen2005fermionic,tranter2018comparison}with the spin-orbital ordering
	\begin{equation}
		\text{qubit } 2 p     \leftrightarrow (p, \uparrow) , \qquad
		\text{qubit } 2 p + 1 \leftrightarrow (p, \downarrow) .
		\label{eq:jw_ordering}
	\end{equation}
	Under this ordering,
	\begin{equation}
		\begin{aligned}
			\hat{a}_q
			= \tfrac{1}{2} \bigl( \hat{X}_q - \hat{X}_q \hat{Z}_q \bigr)
			\prod_{k=0}^{q-1} \hat{Z}_k ,
			\\
			\hat{a}^\dagger_q
			= \tfrac{1}{2} \bigl( \hat{X}_q + \hat{X}_q \hat{Z}_q \bigr)
			\prod_{k=0}^{q-1} \hat{Z}_k ,
			\label{eq:jw_fermion}
		\end{aligned}
	\end{equation}
	and the Hamiltonian becomes a real-valued Pauli sum,
	$\hat{H} = \sum_\alpha c_\alpha \hat{P}_\alpha$ with $\hat{P}_\alpha \in \{\hat{I}, \hat{X}, \hat{Y}, \hat{Z}\}^{\otimes n}$.
	The spin-resolved particle number operators are given by:
	\begin{equation}
		\hat{N}_\uparrow = \sum_{p=0}^{M-1} \frac{1 - Z_{2p}}{2},
		\qquad
		\hat{N}_\downarrow = \sum_{p=0}^{M-1} \frac{1 - Z_{2p+1}}{2}.
	\end{equation}
	The expectation values of these quantities in a state yield the average occupation of the spin-up and spin-down sectors in that state, respectively. 
	\par
	\subsubsection{Parameters, operators and the initial states}
	We shall work with $M=4$ spatial orbitals, with $t=1$ and $U=4$, corresponding to the strong interaction regime. 
	In this illustration, we employ QUEST and ADAPT-VQE for the following two sets of targets:
	\begin{equation}
		\begin{aligned}
			\langle \psi |\hat{H}|\psi\rangle=3, \quad \langle \psi |\hat{N}_\uparrow|\psi\rangle=4, \text{ and }\psi |\hat{N}_\downarrow|\psi\rangle=2\\
			\langle \psi |\hat{H}|\psi\rangle=9, \quad \langle \psi |\hat{N}_\uparrow|\psi\rangle=2, \text{ and }\psi |\hat{N}_\downarrow|\psi\rangle=2
		\end{aligned}
		\label{eq:hubbard_targets}
	\end{equation}
	As for the initial state, we carry out the simulations at five different initial states, listed in Table \ref{tab:initial_states_basis}. The expectation values of energy, up-spin and down-spin number operators are given in Table \ref{tab:initial_states}. Both QUEST and ADAPT-VQE simulations were carried out with $\mathcal{P}_2$ pool. 
	\begin{table*}
		\caption{Fock-space and computational-basis representations of the initial states used in this illustrative example. The generation column expresses each state as a product of fermionic creation operators acting on the Fock vacuum $\ket{\mathrm{vac}}$. Here $\uparrow$, $\downarrow$, $\uparrow\downarrow$, and $\circ$ denote spin-up, spin-down, double occupancy, and empty site, respectively. We adopt the Jordan--Wigner ordering in Eq.~(\ref{eq:jw_ordering}), and represent computational basis states using a binary integer encoding with qubit $0$ as the least-significant (rightmost) bit. Accordingly, each site $p$ is represented by the adjacent qubit pair $(q_{2p+1}, q_{2p}) = (\downarrow, \uparrow)$, so that a two-bit group $b_1 b_0$ encodes $(n_{p\downarrow}, n_{p\uparrow})$. The notation $(\cdots)^{M/2}$ denotes repetition of the enclosed pattern $M/2$ times.}
		\label{tab:initial_states_basis}
		\begin{ruledtabular}
			\begin{tabular}{c l l l l}
				\# & Name & Generation on $\ket{\mathrm{vac}}$
				& Spin configuration & Computational-basis ket \\
				\hline
				
				1 & Doubly-occupied
				& $\displaystyle \prod_{p=0}^{M/2 - 1}
				\hat{a}^\dagger_{p\uparrow}\, \hat{a}^\dagger_{p\downarrow}\,
				\ket{\mathrm{vac}}$
				& $\underbrace{\uparrow\downarrow\,\cdots\,\uparrow\downarrow}_{M/2}\;
				\underbrace{\circ\,\cdots\,\circ}_{M/2}$
				& $\bigl| (11)^{M/2}\, (00)^{M/2} \bigr\rangle$ \\[8pt]
				
				2 & N\'eel
				& $\displaystyle \prod_{p=0}^{M-1}
				\hat{a}^\dagger_{p, \sigma_p}\, \ket{\mathrm{vac}},
				\quad \sigma_p =
				\begin{cases}
					\uparrow & p \text{ even} \\
					\downarrow & p \text{ odd}
				\end{cases}$
				& $\uparrow\,\downarrow\,\uparrow\,\downarrow\,\cdots$
				& $\bigl| (01\,10)^{M/2} \bigr\rangle$ \\[8pt]
				
				3 & Anti-N\'eel
				& $\displaystyle \prod_{p=0}^{M-1}
				\hat{a}^\dagger_{p, \bar{\sigma}_p}\, \ket{\mathrm{vac}},
				\quad \bar{\sigma}_p =
				\begin{cases}
					\downarrow & p \text{ even} \\
					\uparrow & p \text{ odd}
				\end{cases}$
				& $\downarrow\,\uparrow\,\downarrow\,\uparrow\,\cdots$
				& $\bigl| (10\,01)^{M/2} \bigr\rangle$ \\[8pt]
				
				4 & CDW
				& $\displaystyle \prod_{p=0}^{M/2-1}
				\hat{a}^\dagger_{2p, \uparrow}\, \hat{a}^\dagger_{2p, \downarrow}\,
				\ket{\mathrm{vac}}$
				& $\uparrow\downarrow\,\circ\,\uparrow\downarrow\,\circ\,\cdots$
				& $\bigl| (11\,00)^{M/2} \bigr\rangle$ \\[8pt]
				
				5 & Fully polarized
				& $\displaystyle \prod_{p=0}^{M-1}
				\hat{a}^\dagger_{p\uparrow}\, \ket{\mathrm{vac}}$
				& $\uparrow\,\uparrow\,\uparrow\,\cdots$
				& $\bigl| (01)^{M} \bigr\rangle$ \\
				
			\end{tabular}
		\end{ruledtabular}
	\end{table*}
	
	\begin{table}
		\caption{Initial values of the expectation values of the target operators in the QUEST simulation of Quantum State Engineering for the Hubbard chain at half-filling ($N_e = M$ electrons on $M$ sites, $M$ even). Columns list the initial-state label and the expectation values of the three target operators $\hat{H}$, $\hat{N}_\uparrow$, and $\hat{N}_\downarrow$ on the five initial states. The hopping contribution to $\langle \hat{H} \rangle$ vanishes on every state shown, because each hopping term either annihilates the initial state or maps it to an orthogonal computational basis state. The Fock-space and computational-basis representations are given in Table~\ref{tab:initial_states_basis}.}
		\label{tab:initial_states}
		\begin{ruledtabular}
			\begin{tabular}{c l c c c}
				\# & Name & $\langle \hat{H} \rangle$ & $\langle \hat{N}_\uparrow \rangle$ & $\langle \hat{N}_\downarrow \rangle$ \\
				\hline
				1. & Doubly-occupied & $(M/2)\,U$ & $M/2$ & $M/2$ \\
				2. & N\'eel          & $0$        & $M/2$ & $M/2$ \\
				3. & Anti-N\'eel     & $0$        & $M/2$ & $M/2$ \\
				4. & CDW             & $(M/2)\,U$ & $M/2$ & $M/2$ \\
				5. & Fully polarized   & $0$        & $M$   & $0$   \\
			\end{tabular}
		\end{ruledtabular}
	\end{table}
	\subsubsection{Results}
\begin{figure}[h]
		\includegraphics[width=\columnwidth]{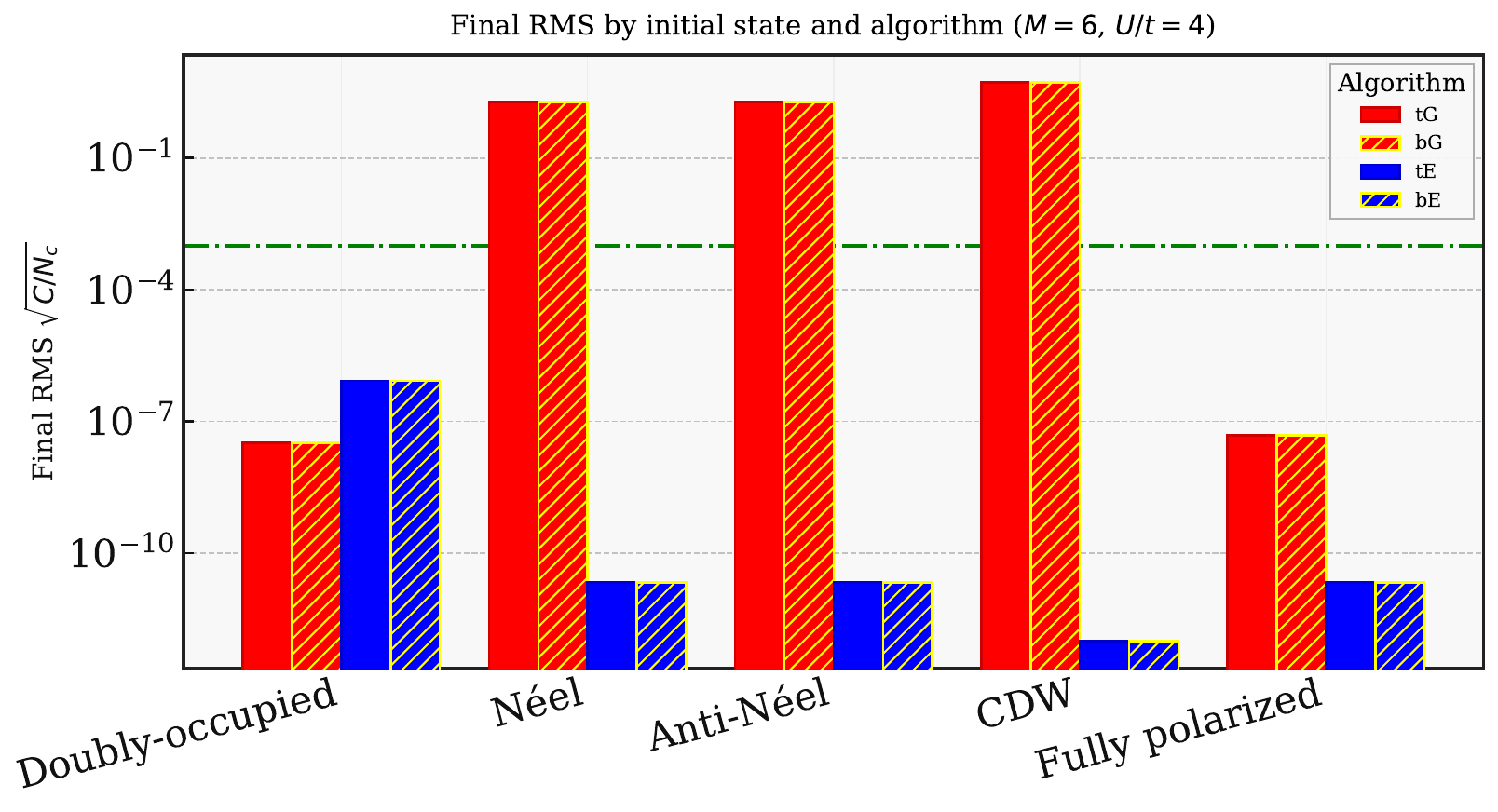}
		\caption{RMS cost per constraint at  termination, for the four implementations of QUEST, for the five intial conditions and for the first set of the three constraints of Eq. (\ref{eq:hubbard_targets}). } 
		\label{fig:Hubbard_Sweep_cost_bars}
	\end{figure}

\begin{figure}[h]
		\includegraphics[width=\columnwidth]{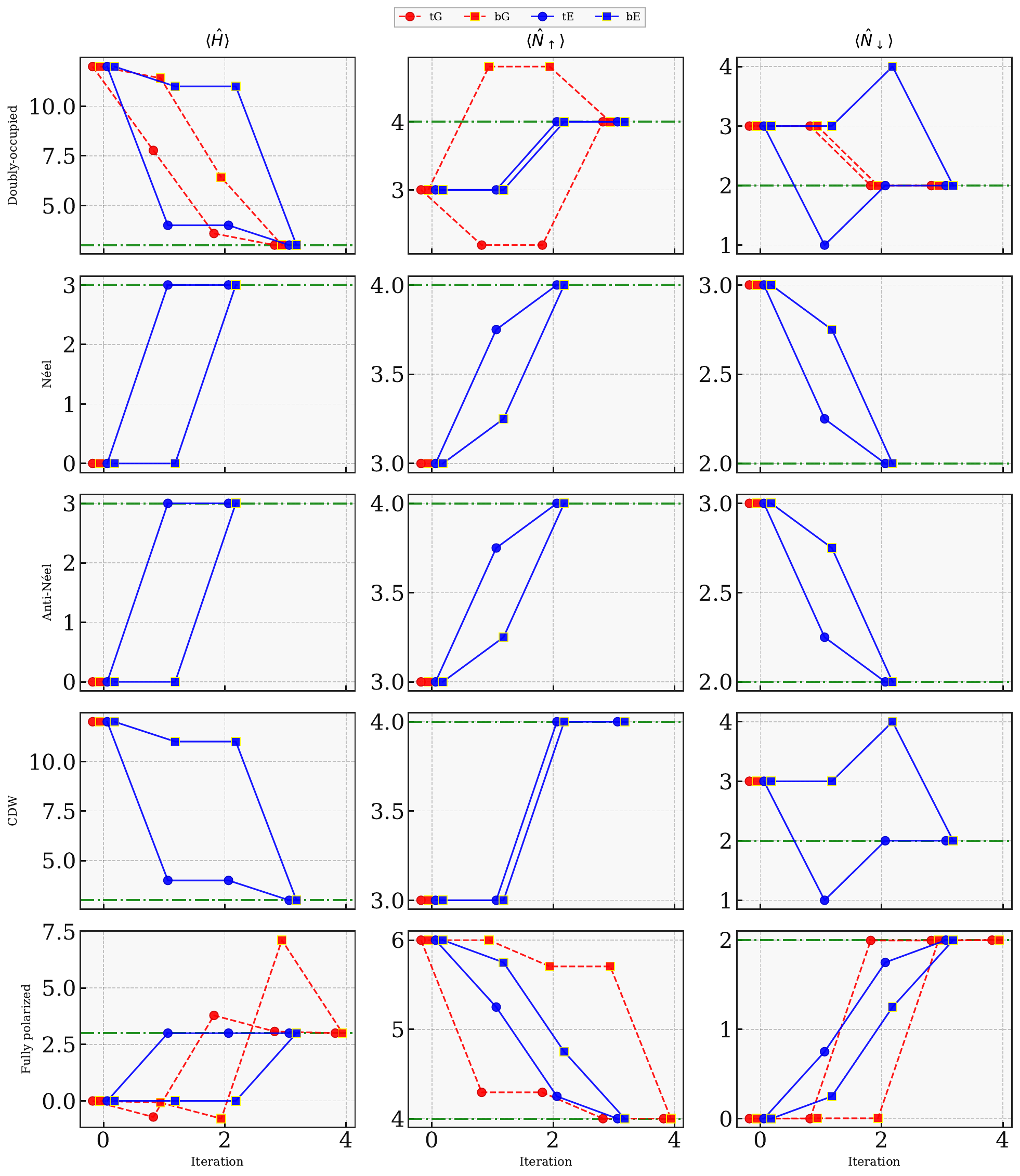}

		\caption{Variation of the expectation values of the three  operators $\hat{H}, \hat{N}_\uparrow$ and $\hat{N}_\downarrow$, as a function of iteration, for the five different initial states, with four implementations of QUEST, with the targets given by first of Eq. (\ref{eq:hubbard_targets})}. 
		\label{fig:Hubbard_Sweep_Grid}
	\end{figure}

	Figs. (\ref{fig:Hubbard_Sweep_cost_bars} and \ref{fig:Hubbard_Sweep_Grid}), show the results corresponding to the first set of targets of Eq. (\ref{eq:hubbard_targets}), starting from the five initial conditions listed in Table . Fig. (\ref{fig:Hubbard_Sweep_cost_bars}) shows the RMS cost per constraint at the termination for the four runs starting from the five initial conditions. 
	Evidently, the gradient based methods (denoted by red bars in this figure) it achieves an RMS cost per constraint value of less than $10^{-7}$ only in two cases. They perform poorly in the N\'eel, Anti-N\'eel and CDW initial conditions. The exact versions QUEST-bE and QUEST-tE (denoted by blue bars in this figure), on the other-hand, converge to an RMS value of better than the set value in all the fives cases. 
	To understand the failure of the gradient implementations of QUEST, we plot the three expectation values are as a function of iteration, in all the fives cases, making a total of $15$ panels, as shown in Fig.  (\ref{fig:Hubbard_Sweep_Grid}),. From this figure it is evident that in the three cases, N\'eel, Anti-N\'eel and CDW, both the gradient-based implementations fail at the first iteration itself. That is, they fail to find any cost-reducing Pauli at the initial state itself.

\begin{figure}[h]
		\includegraphics[width=\columnwidth]{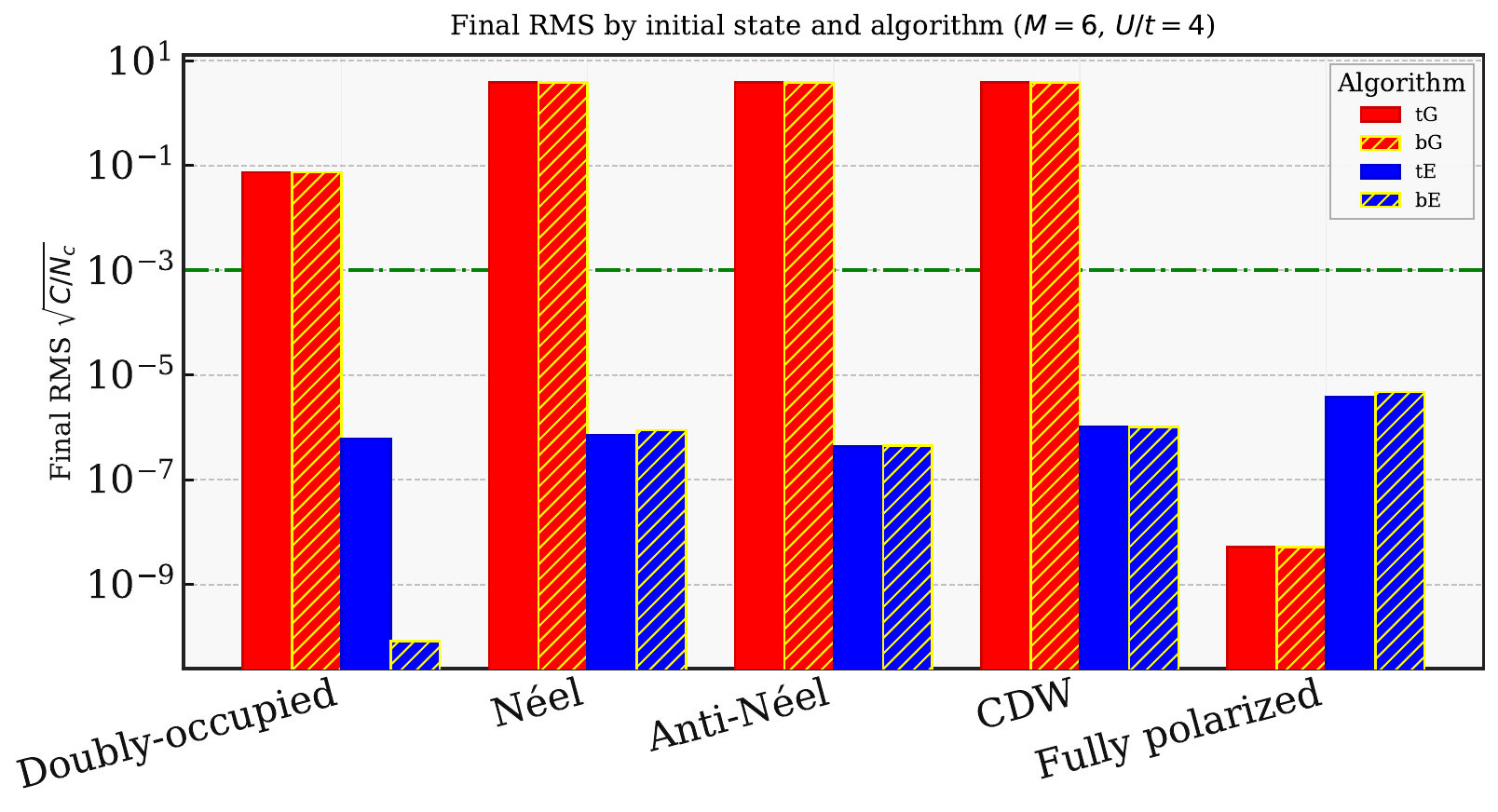}
		\caption{RMS cost per constraint at  termination, for the four implementations of QUEST, for the five intial conditions and for the second set of the three constraints of Eq. (\ref{eq:hubbard_targets}). } 
		\label{fig:Hubbard_Sweep_cost_bars2}
	\end{figure}

\begin{figure}[h]
		\includegraphics[width=\columnwidth]{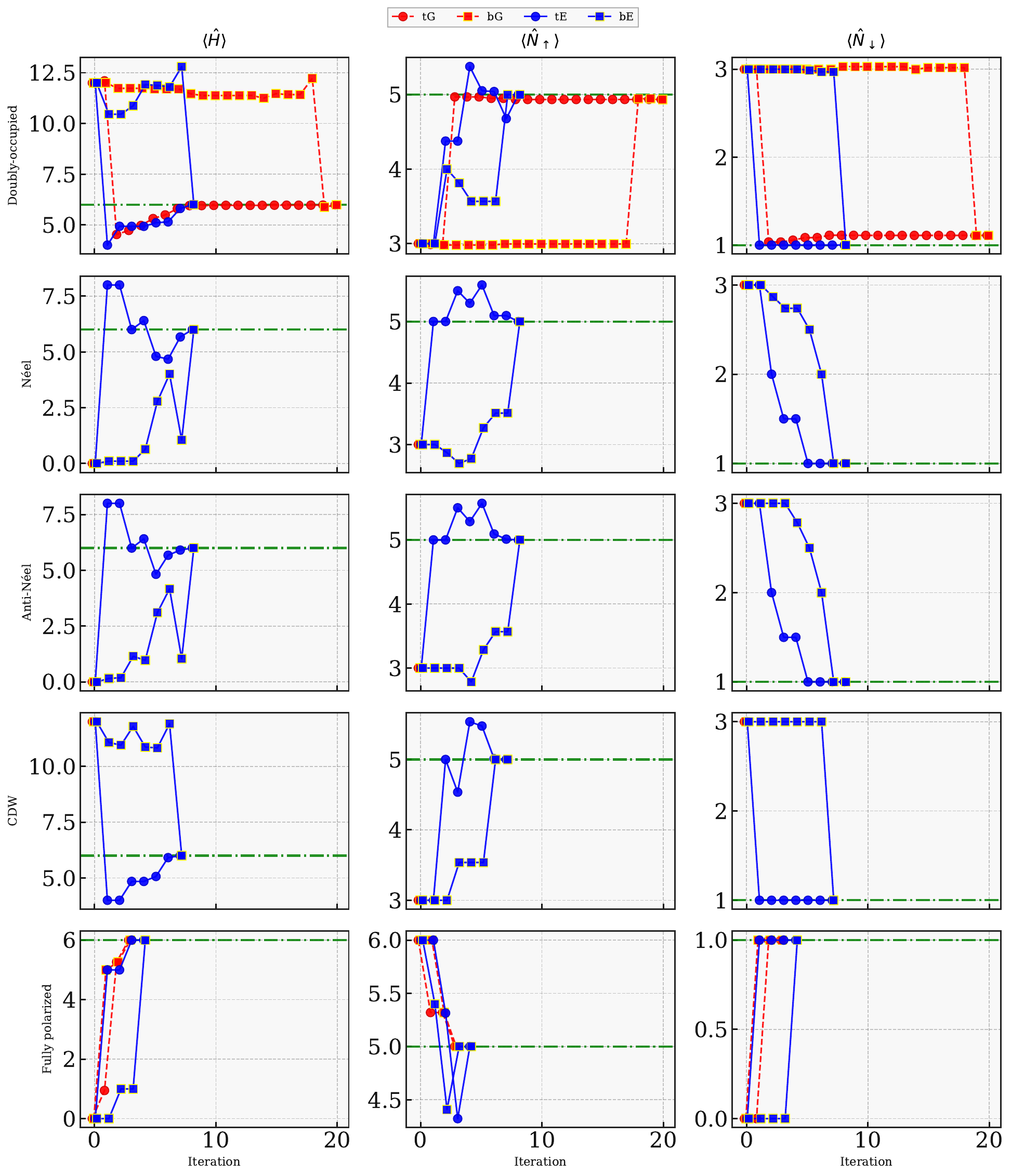}
		
		\caption{Variation of the expectation values of the three constraints $\hat{H}, \hat{N}_\uparrow$ and $\hat{N}_\downarrow$, as a function of iteration, for the five different initial states, with four implementations of QUEST, corresponding to the second set of constraints of Eq. (\ref{eq:hubbard_targets}.) } 
		\label{fig:Hubbard_Sweep_Grid_2}
	\end{figure}

	Figs. (\ref{fig:Hubbard_Sweep_cost_bars2} and \ref{fig:Hubbard_Sweep_Grid_2}), show the results corresponding to the second set of targets of Eq. (\ref{eq:hubbard_targets}). In this case too, QUEST-tE and QUEST-bE succeed in achieving the desired RMS cost per constraint of less than $10^{-3}$ in all the five cases. As in the previous case, the gradient-based implementations stall at the very first iterations, having failed to have found any suitable Pauli. In the fully polarized case, however, the gradient-based implementations succeed in doing better than the non-gradient based implementations. 
	
	\label{subsec:results}
	
	\subsection{A construction of QUEST-tG stalling configurations}
	\label{sec:stalling_construction}
	
	As an illustration of a failure mode unqiue to gradient-based QUEST implementations, we now construct a expectation value targeting problem where the two gradient-based QUEST implementations fail at the initial state itself. That is, the gradients of the cost vanish for all the Paulis $P\in\mathcal{P}$ of all weights, although the cost is far from $0$. Indeed, given any $n$-qubit initial state $|\chi\rangle$, a two-constraint expectation-value targeting problem can be explicitly constructed for which the cost gradient vanishes identically at $|\chi\rangle$ for every Pauli string $P\in\mathcal{P}$. The construction is contrived, in that the operators, targets and initial states are hand-picked to produce algebraic cancellation across the two constraints, but still it demonstrates that gradient based implementations have structural failure modes that can't be resolved with pool enlargement.
	\label{sec:universal_stalling}
	The construction proceeds as follows. Let $\hat{O}_1$  and $\hat{O}_2$ be two Hermitian operators such that the state $\ket{\chi}$ is an eigenstate of the difference operator $\hat{O}_2-\hat{O}_1$.
	Define the targets as
	\begin{equation}
		\tau_1=\langle\chi|\hat{O}_1|\chi\rangle - \delta, \text{ and }\tau_2=\langle\chi|\hat{O}_2|\chi\rangle + \delta, 
	\end{equation} 
	where $\delta \in \mathbb{R}$ is a parameter to ensure that $\tau_1$ and $\tau_2$ are within the spectral limits of $\hat{O}_1$ and $\hat{O}_2$ respectively.
	With the targets so chosen, $\langle\chi|\hat{O}_1|\chi\rangle 
	- \tau_1 = \delta$ and $\langle\chi|\hat{O}_2|\chi\rangle - \tau_2 = -\delta$, so Eq.~\eqref{eq:cost_gradient} evaluated at 
	$|\chi\rangle$ and any $P$ becomes proportional to $\langle\chi|[\hat{O}_1 - 
	\hat{O}_2,P]|\chi\rangle$. Since $|\chi\rangle$ is an eigenstate of $\hat{O}_2 - \hat{O}_1$, this commutator vanishes identically. The gradient is therefore zero for every $P$ in the full Pauli pool, indicating that QUEST-tG and QUEST-bG will terminate at the initial state $\ket{\chi}$ itself. 
	
	As an illustration, consider the transverse-field Ising Hamiltonian under periodic boundary conditions, Eq. (\ref{eq:H_TFIM}), and consider the normalized transverse magnetization operator:
	\begin{equation}
		M_X = \frac{1}{n} \sum_{i=0}^{n-1} X_i.
		\label{eq:MX}
	\end{equation}
	Take the two operators to be $\hat{O}_1 = -nhM_X$, $\hat{O}_2=\hat{H}_{\mathrm{TFIM}}$,
	the initial state to be $|\chi\rangle = |0\rangle^{\otimes n}$ and set the targets as $\tau_1 = -\delta$ and $\tau_2 = -Jn+\delta$ for some  $\delta\in\mathbb{R}$. The operators $\hat{O}_1$ and $\hat{O}_2$ are chosen such that $|\chi\rangle$ is an eigenstate of $\hat{O}_2-\hat{O}_1$.

	QUEST-tG selects pool elements by the magnitude of $|dC/d\theta_P|_{\theta_P = 0}$, a purely local quantity at $\theta_P = 0$. When the cost function admits an algebraic cancellation of gradient contributions across observables  at the initial state, this selection criterion fails completely. 
	
	The other two implementations of QUEST samples $C(\theta)$ at multiple finite angles and reconstruct the full trigonometric landscape. They are therefore immune to this cancellation, and recover the  optimal angle in closed form independently of the slope at the origin.
	\label{sec:Results}
	\clearpage
	\section{Future Directions}
	\label{sec:Future_Directions}
	The present formulation opens several directions for further investigation. At a foundational level, it is of interest to characterize the set of expectation values attainable by pure states and to derive feasibility conditions on the target constraints, thereby clarifying the geometry of the quantum-representable region in operator space. This also connects to the broader question addressed in the quantum marginal problem literature  discussed in the Introduction section, now applied to arbitrary (non-local) operator sets rather than reduced density matrices.
	
	On the algorithmic side, establishing convergence properties of QUEST and quantifying its robustness under finite sampling noise remain important open problems, particularly in relation to the reconstruction of the underlying trigonometric cost landscape. QUEST was presented interms of four implementations, but they differ only in the selection of the Pauli from the pool and the location where it has to be inserted. These choices need not be frozen once for all iterations. For instance, one could choose an implementation depending on the measurement outcomes and the cost reduction in the previous iteration. The two Pauli selection stratagies can also be mixed. Using the gradient information for screening the Pauli strings, and then selecting the best among them with respect to the exact cost reduction they offer can be one such strategy. In fact, in the current implementations, the choice of the Paulis and the location at an iteration does not depend on the Paulis already present in the Pauli path. An efficient strategy to reduce oracle calls is to consider the existing Pauli path when searching for the next Pauli to be inserted. For instance, in the QUEST-bG and QUEST-bE implementations, it is possible to reduce the oracle calls by half, by noting that the cost reducing possible by inserting a Pauli $P$ at the $(j-1)^{\text{th}}$ position is the same as that at the $j^{\text{th}}$ position, if $P$ commutes with the Pauli $P_{j}$ in the current Pauli path.
	
	We have employed an Oracle model for expectation-value estimation, as this abstraction isolates the constraint structure of the problem from implementation-specific details such as observable decomposition and measurement strategies. Such considerations can be incorporated subsequently without altering the core framework. Further refinements may be achieved by incorporating problem-specific structure, such as locality, correlations, and commutation relations among observables. The selection of the Pauli pool can also be made depending on the structure of the problem, and the initial state. The structure of the problem, and the current Pauli path together might also offer opportunities to screen out some Pauli strings and/or some locations from consideration.\\
	Since a majority of the Oracle calls are expended on the selecting the Pauli to be inserted, a better heuristic for selecting or at-least rejecting a Pauli based on some partial information about the cost reduction it affords could be devised. 
	
	In the QUEST implementations presented here, the weight vector $\textbf{w}$ is fixed throughout the QUEST algorithm run. However, the weights $\textbf{w}$ can also be updated between the iterations $\textbf{w}_{k}\rightarrow \textbf{w}_{k+1}$ based on the cost reduction and operator expectation values achieved in the $k^{th}$ iteration. This might give QUEST  flexibility to focus on the most important constraints at each iteration. 
	\section{Conclusion}
	\label{sec:Conclusion}
	In conclusion, we have introduced a formulation of quantum state engineering in which the objective is to prepare a pure multi-qubit state that satisfies a prescribed set of expectation values for a given collection of Hermitian operators. We proposed an algorithmic framework, QUEST, as one concrete approach to solving this problem. An efficient implementation of QUEST accomplishes this task by analytically determining Pauli rotation angles and adaptively inserting Pauli elements at effective locations. This formulation enables the direct enforcement of expectation-value constraints without recourse to variational energy minimization and suggests a route toward efficient use of quantum resources, particularly in the NISQ regime.
	
	We have demonstrated the practical utility of QUEST in solving this class of problems through three distinct examples. The results indicate that QUEST, even with a limited Pauli string pool, is capable of realizing a broad range of constraints without incorporating explicit problem-specific structure into the algorithmic flow.

	By recasting state preparation as expectation-value feasibility, this work emphasizes a viewpoint in which quantum states are characterized directly through observable constraints rather than through a distinguished Hamiltonian. This perspective provides a constructive route to state preparation and highlights a broader set of states defined by their expectation values. Extending this framework to mixed states and quantum channels may offer a unified approach to state and process engineering based on observable constraints.
	\par
	\begin{acknowledgments}
		AM acknowledges the support of Indira Gandhi Centre for Atomic Research and the Homi Bhabha National Institute (HBNI), Mumbai, through the award of a Junior Research Fellowship. 
	\end{acknowledgments}
	\appendix
	\section{Sinusoidal variation of expectation values}
	Here we show that the quantities $O^{alt}(\vec{\theta},\vec{P};\,l;\,\theta)$ and $O^{ins}(\vec{\theta},\vec{P};\,l;\,\theta,\,P)$ defined in Eq. (\ref{eq:operator_expectation_value_alt}), vary sinusoidally with the angle $\theta$.\par
	Let $\ket{\psi}$ be an arbitrary $n-$qubit state and $\hat{O}$ be an arbitrary Hermitian operator. Consider the quantity $\langle \psi \vert e^{-i\theta P}\hat{O}e^{i\theta P} \vert \psi \rangle$. Expanding $e^{i\theta P}= \cos\theta \hat{I}+i\sin\theta P$, we have :
	\begin{equation}
		\begin{aligned}
			\langle \psi \vert e^{-i\theta P}\hat{O}e^{i\theta P} \vert \psi \rangle= &\frac{1}{2} \big(\langle \psi \vert \hat{O}+P\hat{O}P\vert \psi \rangle +\\
			&\cos2\theta\langle \psi \vert \hat{O}-P\hat{O}P\vert \psi \rangle +\\
			&\sin2\theta\langle \psi \vert i[\hat{O},P]\vert \psi \rangle\big).
		\end{aligned}
		\label{eq:A_altered_Exp}
	\end{equation}
	If the state $\ket{\psi}$ is taken as $\ket{\psi}=\hat{U}(\vec{\theta},\vec{P})\ket{\chi}$, where $\ket{\chi}$ is some initial state, then the quantity $\langle \psi \vert e^{-i\theta P}\hat{O}e^{i\theta P} \vert \psi \rangle$ is equal to insert the Pauli $P$ at the end of the Pauli path $\vec{P}$.
	\par
	The three expectation values defined in Eq. (\ref{eq:operator_expectation_value_alt}) in terms of the dressed operator is
	\begin{equation}
		\begin{aligned}
			O^{alt}(\vec{\theta},\vec{P};\,l;\,\theta)=&\langle \psi_{(l)} \vert e^{-i\theta P_l}\tilde{O}^{(l+1)}e^{i\theta P_l}\vert \psi_{(l)}\rangle,\\
			O^{ins}(\vec{\theta},\vec{P};\,l;\,\theta,\,P)=&\langle \psi_{(l+1)} \vert e^{-i\theta P}\tilde{O}^{(l+1)}e^{i\theta P}\vert \psi_{(l+1)}\rangle,\\
			O^{del}(\vec{\theta},\vec{P};\,l)=&\langle \psi_{(l)} \vert \tilde{O}^{(l+1)}\vert \psi_{(l)}\rangle
		\end{aligned}
	\end{equation}
	Therefore the quantities $O^{alt}(\vec{\theta},\vec{P};\,l;\,\theta)$ and $O^{ins}(\vec{\theta},\vec{P};\,l;\,\theta,\,P)$ are of the form $\langle \psi \vert e^{-i\theta P}\hat{A}e^{i\theta P} \vert \psi \rangle$. Therefore they vary sinusoidally with the angle $\theta$ as given in Eq. (\ref{eq:A_altered_Exp}).
	\bibliographystyle{apsrev4-2}
	\bibliography{quest}
	
\end{document}